\documentclass[useAMS,usenatbib]{mn2e}

\usepackage{graphicx}
\usepackage{amsmath}
\usepackage{amsfonts}
\usepackage{amssymb,latexsym}
\usepackage[english]{babel}
\usepackage{mathcomp}
\usepackage{dsfont}
\usepackage{natbib}
\usepackage[usenames]{color}
\usepackage{array}

\title[Characterisation of the non-Gaussianity of radio and IR point
  sources at CMB frequencies] 
{Characterisation of the non-Gaussianity of radio and IR point sources
  at CMB frequencies}  
\author[F. Lacasa, N. Aghanim, M. Kunz, and M. Frommert]
{F. Lacasa$^{1}$\thanks{E-mail: Fabien.Lacasa@ias.u-psud.fr}, 
N. Aghanim$^{1}$, 
M. Kunz$^{2}$, 
and M. Frommert$^{2}$ \\
$^{1}$ Institut d'Astrophysique Spatiale (IAS), B\^atiment 121, F-91405 Orsay
(France); Universit\'e Paris-Sud 11 and CNRS (UMR 8617) \\
$^{2}$ D\'{e}partement de Physique Th\'{e}orique and Center for
Astroparticle Physics, Universit\'e de Gen\`{e}ve, 24 quai
E. Ansermet, Gen\`{e}ve }

\begin{document}

\date{}

\pagerange{\pageref{firstpage}--\pageref{lastpage}} \pubyear{2011}

\maketitle

\label{firstpage}
 
\begin{abstract}
This study, using publicly available simulations, focuses on
the characterisation of the non-Gaussianity produced by radio point
sources and by infrared (IR) sources in the frequency
range of the cosmic microwave background from 30 to 350 GHz.

We propose a simple prescription to infer the angular bispectrum from the
power spectrum of point sources considering independent
populations of sources, with or without clustering. We test the
accuracy of our prediction using publicly available all-sky
simulations of radio and IR sources and find very good agreement.

We further characterise the configuration dependence and the frequency
behaviour of the IR and radio bispectra. We show that the IR
angular bispectrum peaks for squeezed triangles and that the clustering of IR
sources enhances the bispectrum values by several orders of magnitude
at scales $\ell \sim 100$.  At 150 GHz the bispectrum of IR sources
starts to dominate that of radio sources on large angular scales,
and it dominates over the whole multipole range at 350 GHz.

Finally, we compute the bias on $f_\mathrm{NL}$ induced by radio and
IR sources. We show that the positive bias induced by radio sources is
significantly reduced by masking the sources.  We also show, for the
first time, that the form of the IR bispectrum mimics a primordial
`local' bispectrum $f_\mathrm{NL}$. The IR sources produce a negative
bias which becomes important for Planck-like resolution and at high
frequencies ($\Delta f_\mathrm{NL} \sim -6$ at 277 GHz and $\Delta
f_\mathrm{NL} \sim -$60-70 at 350 GHz). Most of the signal being due
to the clustering of faint IR sources, the bias $\Delta
f_\mathrm{NL}^\mathrm{IR}$ is not reduced by masking sources above a
flux limit and may, in some cases, even be increased due to the
reduction of the shot-noise term.

\end{abstract}



\section{Introduction}

In the last few decades, the Cosmic Microwave Background (CMB) has become
a very successful probe of the early and late universe. The
smallness of the perturbations in the cosmic fluids, and hence in the space-time metric,
allows us to use linear perturbation theory to compute their evolution
efficiently and accurately through Boltzmann codes (\cite{Seljak1996},
\cite{Lewis2000}, \cite{Lesgourgues2011}).

Since COBE \citep{Smoot1992} the measurement of the CMB power spectrum
has been achieved by many experiments and over a wide range of
scales. The most recent CMB data are those of the Seven-Year Wilkinson Microwave Anisotropy Probe (WMAP)
\citep{Larson2011}, Atacama Cosmology Telescope (ACT)
\citep{Das2011} and South Pole Telescope (SPT)
\citep{Keisler2011}. Constraints from all these measurements, combined
with probes of the geometry of the Universe , like Baryonic Acoustic
Oscillations (e.g. \cite{Percival2010}, \cite{Blake2011}), type Ia
Supernovae (e.g. \cite{Astier2006}, \cite{Hicken2009}, \cite{Guy2010})
and Hubble constant measurements (e.g. \cite{Freedman2001},
\cite{Riess2009}, \cite{Freedman2009}), give a converging view of our
Universe and have lead to the establishment of a `standard model' of
cosmology (e.g. \cite{Larson2011}) known as $\Lambda$CDM. In this
model, the universe is flat, dominated by a cold dark matter component
and a `dark energy' component compatible with a cosmological
constant. The present constraints suggest that CMB anisotropies are a
realisation of a primordial random process that generated the initial
perturbations from quantum fluctuations which were then stretched to
macroscopic scales by inflation (e.g. \cite{Starobinski1979},
\cite{Guth1981}, \cite{Liddle2000}, see \cite{Bassett2006} and
\cite{Linde2008} for reviews).

The microwave sky is however not made of CMB primary signal alone, it
consists also of secondary anisotropies such as those associated with
the Integrated Sachs-Wolfe effect -- arising from the time-variable
  metric perturbations \citep{SW1967} --, those arising from
the Sunyaev-Zel'dovich (SZ) effect (inverse Compton scattering) in the
direction of galaxy clusters \citep{Sunyaev1972}, those due to the
Doppler effect from moving structures (e.g. kinetic SZ effect from
clusters and inhomogeneous reionisation), see \cite{Aghanim2008} for a
review on secondary anisotropies. In addition, there are other
astrophysical components in the microwave domain such as the dust,
synchrotron and free-free emissions from our Galaxy
\citep{Planck-Collaboration-Dust}, the emission from radio sources
that dominate at lower frequencies but contribute significantly at
microwave frequencies \citep{deZotti2005}, and the emission from dusty
star-forming galaxies emitting mainly in the Infra-Red (IR) domain
\citep{Low1968}.

In this study, we will focus only on the characterisation of
the extra-galactic point sources, namely the radio sources and the IR
dusty galaxies. They contribute notably to the power spectrum at CMB
frequencies and start dominating over the CMB at about $\ell \sim$
2000 when the CMB signal plummets.  Active Galactic Nuclei (AGN) are
observed as radio sources via their synchrotron emission. They have
been widely studied in the CMB context especially at low frequencies
$\nu \leq 90$ GHz (e.g. \cite{deZotti2005}, \cite{Boughn2008},
\cite{Sajina2011}). They affect mostly the lower end of frequencies
observed by CMB experiments. Their largest impact was thought to be in
the frequency bands from 30 to 90 GHz but the recent Planck results
\citep{Planck-Collaboration-ERCSC, Planck-Collaboration-radioSED} show
that radio sources are detected at frequencies as high as 217 GHz. At
the CMB frequencies, radio sources do not cluster (\cite{Toffolatti1998} 
\cite{Gonzalez2005}) and 
thus exhibit a flat power-spectrum (see Appendix \ref{appendix:shot}).

Star-forming galaxies are observed as IR sources via the thermal
emission from dust heated by the ultra-violet emission of young
stars. These IR sources contribute to the signal observed by CMB
experiments at frequencies higher than 150 GHz, thus they are
  particularly relevant for the most recent CMB experiments, e.g. SPT, ACT and
Planck High Frequency Instrument observations. The cumulated emission
from the IR sources, the Cosmic Infrared Background (CIB), was first
discovered by \cite{Puget1996}, and its anisotropies were first
characterised by \cite{Lagache2000} and \cite{Matsuhara2000}. Many
other observations were possible in the last decade in the IR and
submm domain (\cite{Lagache2007}, \cite{Viero2009}, \cite{Hall2010},
\cite{Amblard2011}, \cite{Planck-Collaboration-CIB}). In particular,
the latest constraints of the CIB from Planck showed that its power
spectrum, at frequencies 217-353-545-857 GHz, is well modeled by a
power law $C_\ell^\mathrm{CIB}(\nu) = A(\nu)\times
\left(\frac{\ell}{1000}\right)^n$ with e.g. $A=(104 \pm 4)\times 10^2
\mathrm{Jy}^2/\mathrm{sr}$ and $n=-1.08 \pm 0.06$ at 545 GHz. This
behaviour contrasts with the flat spectrum of radio sources and is due
to the clustering of the IR galaxies and their host dark-matter halos.

These point sources are super-imposed on the primordial fluctuations.
The simplest models of inflation (single-field and slow-roll) predict
a small primordial non-Gaussianity (NG) (\cite{Acquaviva2003},
\cite{Maldacena2003}, \cite{Creminelli2004})  that is sub-dominant to
the NG induced by the 
non-linear evolution of the perturbations, a contribution that is
necessarily always present. More complex inflationary models,
e.g. multi-field scenarios, may predict larger NG \citep{Byrnes2010},
to the point of being detectable. A simple and yet powerful probe of
non-Gaussianity is the three-point function in harmonic space, the
angular bispectrum (see section \ref{sect:3pNGe} for more details), which is
defined as a function of a multipole triplet ($\ell_1,\ell_2,\ell_3$).
The angular bispectrum vanishes for a Gaussian field like all odd-order
moments. Besides the bispectrum, connected even-order moments may also
be used to probe non-Gaussianity, and the 4-point function or
trispectrum has indeed been a focus of interest \citep{Kunz:2001ym},
especially for lensing studies \citep{Cooray2002}.

There are many different models of inflation, and they often predict
very similar power spectra that are close to being scale
invariant. For models that lead to a measurable bispectrum, however,
this degeneracy can often be broken by studying the dependence of the
bispectrum amplitude on ($\ell_1,\ell_2,\ell_3$), e.g. a large signal
for squeezed triangles is indicative of slow-roll multi-field
inflation models, equilateral triangles are enhanced for models with
non-canonical kinetic terms, and folded triangles for non-standard
vacuum initial conditions (e.g. \cite{Bartolo2004},
\cite{Renaux-Petel2009}, \cite{Lewis2011}).  The most studied and
constrained form of non-Gaussianity is the local ansatz parametrised
by a factor $f_\mathrm{NL}$, and predicted by several inflation
models. Current constraints on local non-Gaussianity show that the CMB
is consistent with being Gaussian at the 95$\%$
C.L. \citep{Komatsu2011}, and constraints on other shapes all show
consistency with Gaussianity as well.

Given the current limits on primordial NG, astrophysical signals are
the dominant contribution to non-Gaussianity. While Galactic emission
and resolved sources may be accounted for by masking, unresolved
sources and residuals have to be characterised. As opposed to
primordial NG, radio sources NG has been detected, and was studied for
the WMAP mission showing that it yields a non-zero flat angular bispectrum
parametrised as $b_\mathrm{src}$ or $b_\mathrm{ps}$. The WMAP5 best
estimate in the Q band is $b_\mathrm{src} = 4.3 \pm 1.3 \mu K^3
\mathrm{sr}^2$\citep{Komatsu2009}. Characterising the NG signal from
astrophysical components and more importantly from extra-galactic
point sources is important for two main reasons: (i) to avoid
mistaking it for a primordial contribution (and to allow the
development of robust methods to isolate primordial NG) and (ii) to
learn more about astrophysical processes, i.e. to go beyond the
description of point sources by their number counts and their power
spectrum.

The study of NG from extra-galactic point sources has been pioneered
by \cite{Argueso2003}, focusing mostly on the radio sources and
including clustering. They showed that the point-source angular bispectrum is
mostly flat at WMAP frequencies and dominates the CMB bispectrum in
most configurations. \cite{Babich2008} investigated the bias on the
$f_\mathrm{NL}$ estimator induced by radio sources, considering the
modulation of their number density with ISW, of their magnification
with lensing and of the flux cut-off with selection
effects. \cite{Serra2008} studied the bias on $f_\mathrm{NL}$ due to
radio sources, SZ-lensing and ISW-lensing bispectra, investigating the
dependence of this bias with the resolution scale. Finally
\cite{Munshi2009} defined skew-spectra for cross-correlation analysis, derived
estimators for the skew-spectra and applied it to secondary anisotropies.

In this paper we study the non-Gaussianity produced by infrared and
radio point sources in the frequency range of the CMB from 30 to 350
GHz, based on numerical simulations by \cite{Sehgal2010}. We
investigate the frequency and configuration dependence of the
angular bispectrum. We particularly focus on the non-Gaussianity from IR
sources and their clustering term. We restrict the study to the
simplest case of full sky maps without masks. Furthermore, we do not
take into account noise and beam effects. Statistical isotropy of all
fields considered will be assumed throughout this article. The case of
noisy, partially masked maps will be addressed in future studies.

In section \ref{sect:3pNGe} we review the estimator of the (binned)
angular bispectrum and $f_\mathrm{NL}$ and develop a parametrisation of the
bispectrum to display and visualise it efficiently. In section
\ref{sect:prescription}, we develop a prescription to infer the
bispectrum from the power spectrum for clustered sources and for
different populations. In section \ref{sect:resultsehgal} we use
publicly available full-sky simulations of radio and infrared sources to
compute and characterise their bispectrum at CMB frequencies and we
compare them to the predictions from the prescription. We
examine the configuration dependence of the point-source bispectra
and study the bias they induce on the estimation of the primordial
local non-Gaussianity in 
section \ref{sect:ngcsqce}. We finally conclude and discuss our results
in section \ref{sect:concl}.


\section{Three-point NG estimators}\label{sect:3pNGe}

%
\subsection{Full-sky angular bispectrum estimator}
Given a full-sky map of the temperature fluctuations $\Delta
T(\mathbf{n})$ of some signal, it can be decomposed in the spherical
harmonic basis 
\begin{equation}
a_{\ell m} = \int \mathrm{d}^2\mathbf{n} \; Y^*_{\ell m}(\mathbf{n}) \; \Delta
T(\mathbf{n})
\end{equation}
with the usual orthonormal spherical harmonics $Y_{\ell m}$
\begin{equation*}
\int \mathrm{d}^2 \mathbf{n} \; Y_{\ell m}(\mathbf{n}) \; Y^*_{\ell'
  m'}(\mathbf{n}) \, 
= \, \delta_{\ell \ell'} \, \delta_{m m'}.
\end{equation*}
Observational data is pixelised, so that the integral is replaced by a
sum over pixels. We   
will assume that the solid angle of a pixel, $\Omega_{\mathrm{pix}}$,
is a constant,  
which is for example the case for the
HEALPix\footnote{http://healpix.jpl.nasa.gov} pixelisation scheme that
we will adopt for the numerical calculations. In this case we have
that 
\begin{equation}
a_{\ell m} = \sum_{\mathbf{n}_i} \; Y^*_{\ell m}(\mathbf{n}_i) \; \Delta
T(\mathbf{n}_i) \, \Omega_{\mathrm{pix}} \,.
\end{equation}
This discreteness effect will be important e.g. in section
\ref{sect:shot}. 

In order to compute the angular bispectrum, which is the harmonic
transform of the 3-point correlation function, we will resort to
scale-maps as defined by \cite{Spergel1999} and also used by 
\cite{Aghanim:2003fs} and \cite{DeTroia:2003tq},
\begin{equation} 
\label{scalemapfullsky}
T_{\ell}(\mathbf{n}) = \sum_m a_{\ell m} Y_{\ell m}(\mathbf{n}) = \int
\! \mathrm{d}^2\mathbf{n}' \, \Delta T(\mathbf{n}') 
\, P_{\ell}(\mathbf{n}\cdot \mathbf{n}')
\end{equation}
where $P_{\ell}$ is the Legendre polynomial of order $\ell$.
\\The optimal bispectrum estimator is then \citep{Spergel1999}:
\begin{multline} 
\label{eq:estimbisp}
\hat{b}_{\ell_1 \ell_2 \ell_3} = \frac{4\pi}{(2\ell_1+1)(2\ell_2+1)(2\ell_3+1)}
\begin{pmatrix}
\ell_1 & \ell_2 & \ell_3 \\ 0 & 0 & 0
\end{pmatrix}^{-2}
\\ \times \int \mathrm{d}^2\mathbf{n} \; T_{\ell_1}(\mathbf{n}) \,
T_{\ell_2}(\mathbf{n}) \, T_{\ell_3}(\mathbf{n})   
\end{multline}
or it can be written in the form:
\begin{eqnarray}
\label{eq:biseq}
\nonumber\hat{b}_{\ell_1 \ell_2 \ell_3} & = &
\sqrt{\frac{4\pi}{(2\ell_1+1)(2\ell_2+1)(2\ell_3+1)}} 
\begin{pmatrix}
\ell_1 & \ell_2 & \ell_3 \\ 0 & 0 & 0
\end{pmatrix}^{-1} \\ 
&& \!\!\times \!\!\!\! \sum_{m_1,m_2,m_3} \!\!
\begin{pmatrix}
\ell_1 & \ell_2 & \ell_3 \\ m_1 & m_2 & m_3
\end{pmatrix}
a_{\ell_1 m_1} \;\! a_{\ell_2 m_2} \;\! a_{\ell_3 m_3}
\end{eqnarray}
where the expression in brackets represents the Wigner $3j$ symbols.
Equation (\ref{eq:biseq}) is computationally expensive when implemented
at high $\ell$ due to the large number of Wigner symbols to
calculate. Equation (\ref{eq:estimbisp}) still requires a few cpu-days for a
full computation at a Planck-like resolution, Nside=1024 -
2048. Binning the multipoles in $\ell$, as \cite{Bucher2010},
has the advantage of speeding up the computations and smoothing
out the variations due to cosmic variance.

For a given triangle in harmonic space
$(\ell_1,\ell_2,\ell_3)$ the number of independent configurations on
the sphere is:
\begin{equation} \label{eq:ntri}
N_{\ell_1 \ell_2 \ell_3} = \frac{(2\ell_1+1)(2\ell_2+1)(2\ell_3+1)}{4\pi}
\begin{pmatrix}
\ell_1 & \ell_2 & \ell_3 \\ 0 & 0 & 0
\end{pmatrix}^2 
\end{equation}

When multipoles are binned in bins of width $\Delta\ell$ the
expression for the scale-maps (Eq. \ref{scalemapfullsky}) becomes:

\begin{equation}
T_{\mathrm{\Delta\ell}}(\mathbf{n}) = \sum_{\ell \in \mathrm{\Delta\ell}, m}
a_{\ell m} \,Y_{\ell m}(\mathbf{n}) 
\end{equation}
and a binned bispectrum estimator identically weighting triangles is
given by:
\begin{equation}
\hat{b}_{\mathrm{\Delta\ell_1,\Delta\ell_2,\Delta\ell_3}} =
\frac{1}{N_{\Delta}(\mathrm{\Delta\ell_1,\Delta\ell_2,\Delta\ell_3})} 
 \int \mathrm{d}^2\mathbf{n} \, T_{\mathrm{\Delta\ell_1}}(\mathbf{n}) \,
 T_{\mathrm{\Delta\ell_2}}(\mathbf{n}) \, 
 T_{\mathrm{\Delta\ell_3}}(\mathbf{n})  
\end{equation}
where
\begin{equation*}
N_{\Delta}(\mathrm{\Delta\ell_1,\Delta\ell_2,\Delta\ell_3})
=\sum_{\ell_i \in \mathrm{\Delta\ell}_i} N_{\ell_1 \ell_2 \ell_3}  
\end{equation*}

One can easily check that the obtained binned bispectrum estimator is
unbiased for a constant bispectrum and that the bias can be neglected
as long as the bispectrum does not vary significantly within a bin
$\Delta\ell$. In the following, we have chosen $\ell_\mathrm{max}=2048$ and a bin width $\Delta\ell=64$ 
for simplicity and computational speed while retaining enough information on the scale dependence \citep{Bucher2010}.


\subsection{$f_{\mathrm{NL}}$ estimator}\label{sect:fnlestim}
The most studied and constrained form of primordial non-Gaussianity is the local
ansatz, whose amplitude is parametrised by a non-linear coupling constant $f_\mathrm{NL}$: 
\begin{equation}\label{eq:fnldef}
\Phi(x) = \Phi_G(x) + f_\mathrm{NL} \left(\Phi^2_G(x)-
 \langle \Phi^2_G(x) \rangle \right) 
\end{equation}
where $\Phi(x)$ is the Bardeen potential and $\Phi_G(x)$ is a Gaussian
field. This form of NG yields the following CMB angular bispectrum
\citep{Komatsu:2001rj}:  
\begin{equation}\label{eq:cmbispth}
b_{\ell_1 \ell_2 \ell_3}^\mathrm{loc} = \int r^2 \, dr \,
\alpha_{\ell_1}(r) \, \beta_{\ell_2}(r) \, \beta_{\ell_3}(r)
+\mathrm{perm.} 
\end{equation}
with
\begin{eqnarray}
\alpha_\ell(r) & = & \frac{2}{\pi} \int k^2 \mathrm{d}k \,
g_{T,\ell}(k) \, j_\ell(kr)  \\
\beta_\ell(r) & = & \frac{2}{\pi} \int k^2 \mathrm{d}k \, P(k) \, g_{T,\ell}(k)
\, j_\ell(kr) 
\end{eqnarray}
where $g_{T,\ell}$ is the radiation transfer function, which can be
computed with a Boltzmann code, $j_{\ell}$ are the spherical Bessel
functions, and $P(k)\propto k^{n_s-4}$ is the primordial power
spectrum, with a spectral index $n_s$.
\\On large angular scales, the Sachs-Wolfe (SW) effect is the dominant
contribution to the CMB signal. In this regime, the CMB bispectrum
takes the following analytical form: 
\begin{equation}\label{eq:cmbispsw}
b_{\ell_1 \ell_2 \ell_3}^\mathrm{loc} \propto -\left(\frac{1}{\ell_1^2
  \, \ell_2^2} +\frac{1}{\ell_1^2 \, \ell_3^2}+\frac{1}{\ell_2^2 \,
  \ell_3^2}\right) \,.
\end{equation}
This bispectrum is maximal when one of the multipoles is minimal
($\ell_1 \ll \ell_2 \simeq \ell_3$) which is called the squeezed
configuration. 

A commonly used cubic estimator of $f_{\mathrm{NL}}$ has been
developed by \cite{Komatsu2005}. It is much faster than performing the
whole bispectrum analysis and fitting the local bispectrum. In its
original version, the estimator takes into account beam profile and
homogeneous noise, and has been used on WMAP data to yield the current
constraint $-10 < f_\mathrm{NL} < 74$ \citep{Komatsu2011}. The
estimator was then further developed by several authors by adding a
linear term accounting for masking and inhomogeneous noise
\citep{Creminelli2006}. Here, we will only consider noiseless full-sky
maps without beam smoothing so that we can apply the estimator in its
original form.

To build the $f_{\mathrm{NL}}$ estimator we first define the filtered
maps at comoving distance $r$ and direction $\mathbf{n}$: 
\begin{eqnarray}
A(r,\mathbf{n}) & = & \sum_{\ell m} \frac{\alpha_\ell(r)}{C_\ell} \, a_{\ell m}
\, Y_{\ell m}(\mathbf{n}) \\ 
B(r,\mathbf{n}) & = & \sum_{\ell m} \frac{\beta_\ell(r)}{C_\ell} \, a_{\ell m}
\, Y_{\ell m}(\mathbf{n}) 
\end{eqnarray}
where $C_\ell$ is the CMB power spectrum. $B(r,\mathbf{n})$ is then an
estimated map of the primordial potential fluctuations
$\Phi(r\;\!\mathbf{n})$ via Wiener filtering.  

The integral of $A B^2$ permits us to estimate $f_\mathrm{NL}$ as:
\begin{equation}
\hat{f}_\mathrm{NL} \!\! \sum_{\ell_1 \leq \ell_2 \leq \ell_3}
\frac{\left(B^{\mathrm{loc}}_{\ell_1 \ell_2
    \ell_3}\right)^2}{C_{\ell_1} C_{\ell_2} C_{\ell_3}} = \int r^2 \mathrm{d}r
\, \mathrm{d}^2\mathbf{n} \, A(r,\mathbf{n}) \, B^2(r,\mathbf{n}) 
\end{equation}
where $B^{\mathrm{loc}}_{\ell_1 \ell_2 \ell_3}=\sqrt{N_{\ell_1 \ell_2
    \ell_3}} \, b^{\mathrm{loc}}_{\ell_1 \ell_2 \ell_3}$ is the local
bispectrum for $f_\mathrm{NL}=1$, to be compared with the observed value
$B^{\mathrm{obs}}_{\ell_1 \ell_2 \ell_3}$.
\\It can be shown that this estimator takes analytically the form:
\begin{equation}\label{eq:Dfnlanalyt}
\hat{f}_\mathrm{NL} = \frac{
\sum_{\ell_1 \leq \ell_2 \leq \ell_3} \frac{B^{\mathrm{obs}}_{\ell_1
    \ell_2 \ell_3} \, B^{\mathrm{loc}}_{\ell_1 \ell_2
    \ell_3}}{C_{\ell_1} C_{\ell_2} C_{\ell_3}}} 
{\sum_{\ell_1 \leq \ell_2 \leq \ell_3}
  \frac{\left(B^{\mathrm{loc}}_{\ell_1 \ell_2
      \ell_3}\right)^2}{C_{\ell_1} C_{\ell_2} C_{\ell_3}}} 
\end{equation} 
It is near-optimal in the sense that it minimizes the $\chi^2$ for
weak NG (under some assumptions on isosceles and equilateral
triangles). The estimator becomes sub-optimal (e.g. \cite{Elsner2009})
for large enough $f_\mathrm{NL}$, when the variance of the bispectrum
gets $\mathrm{O}\!\left(f_\mathrm{NL}^2\right)$ correction compared to
the weak NG computation with Wick's theorem (see Appendix
\ref{appendix:wngvar}).

%
%
\subsection{Parametrisation of the angular bispectrum}\label{sect:param}

Several ways of visualising the angular bispectrum have been proposed
in the literature, e.g. isosurfaces in the ($\ell_1,\ell_2,\ell_3$) 3D
space by \cite{Fergusson2010}, or slices of constant perimeter in the
orthogonal transverse coordinate ($\ell _{\perp a},\ell _{\perp b}$)
space by \cite{Bucher2010}.

Under the assumption of statistical isotropy, the bispectrum
$b_{\ell_1 \ell_2 \ell_3}$ is invariant under permutations of
$\ell_1$, $\ell_2$ and $\ell_3$, ie. it is a function of the shape and size of
the triangle $(\ell_1, \ell_2, \ell_3)$ only, i.e. independent of its
orientation.  Therefore, we can 
find a parametrisation invariant under permutation of $\ell_1$,
$\ell_2$, and $\ell_3$, which avoids redundancy of
information and allows convenient visualisation and
interpretation of data. Let us denote as
$\overline{(\ell_1,\ell_2,\ell_3)}$ the equivalence class of the
triplet under permutations. 

\begin{figure}
\centering
\includegraphics[width=8.5cm]{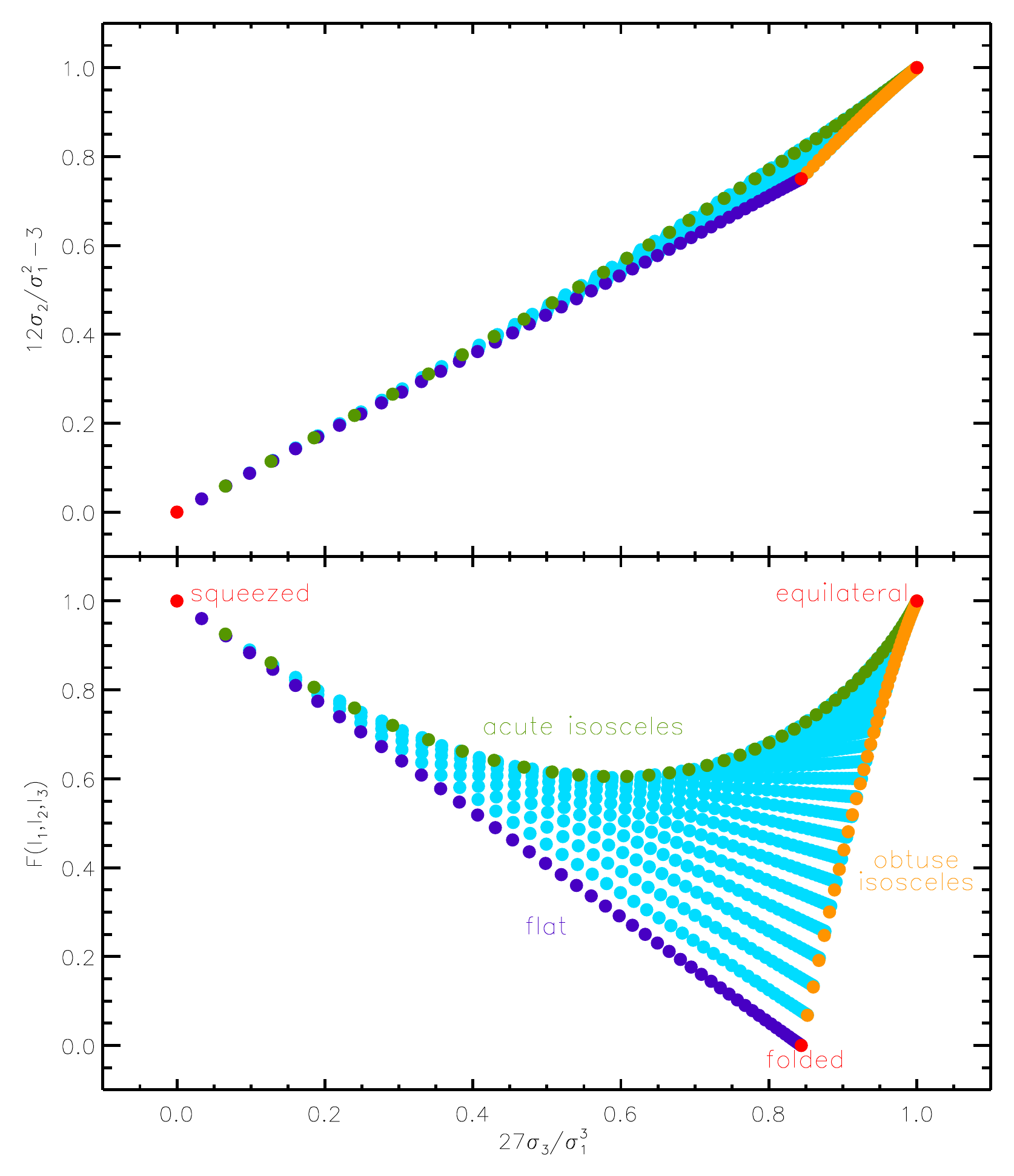}
\caption{Triangles of constant perimeter, $P$, in the parametrisation
  defined by the normalised symmetric polynomials (upper panel), or in
  the parameter space ($F,S$) defined in the text (bottom panel).} 
\label{triparam}
\end{figure}

The elementary symmetric polynomials ensure the invariance under permutations:
\begin{itemize}
\item $\sigma_1 = \ell_1+\ell_2+\ell_3 $
\item $\sigma_2 = \ell_1\ell_2+\ell_1\ell_3+\ell_2\ell_3 $
\item $\sigma_3 = \ell_1\ell_2\ell_3 $
\end{itemize}
Through Cardan's formula, there is a one-to-one correspondence between
$\overline{(\ell_1,\ell_2,\ell_3)}$, defined by the roots of the
polynomial $X^3 - \sigma_1 X^2 + \sigma_2 X - \sigma_3$, and the
triplet $(\sigma_1,\sigma_2,\sigma_3)$. We further define the
scale-invariant parameters $\tilde{\sigma}_2=12\sigma_2/\sigma_1^2-3$
and $\tilde{\sigma}_3=27\sigma_3/\sigma_1^3$ with coefficient chosen
so that $\tilde{\sigma}_2$ and $\tilde{\sigma}_3$ vary in the range
[0,1].  As illustrated in the upper panel of Fig. \ref{triparam}, this
parametrisation does not allow us to discriminate efficiently between
the different triangles.

We find that the parameters noted $(P,F,S)$ and defined as:
\begin{itemize}
\item $P=\sigma_1$
\item $F=32(\tilde{\sigma}_2-\tilde{\sigma}_3)/3+1$
\item $S=\tilde{\sigma}_3$
\end{itemize}
provide a clearer distinction of the triangles as is illustrated in
the bottom panel of Fig. \ref{triparam}. 

\begin{figure}
\centering
\includegraphics[width=8.5cm]{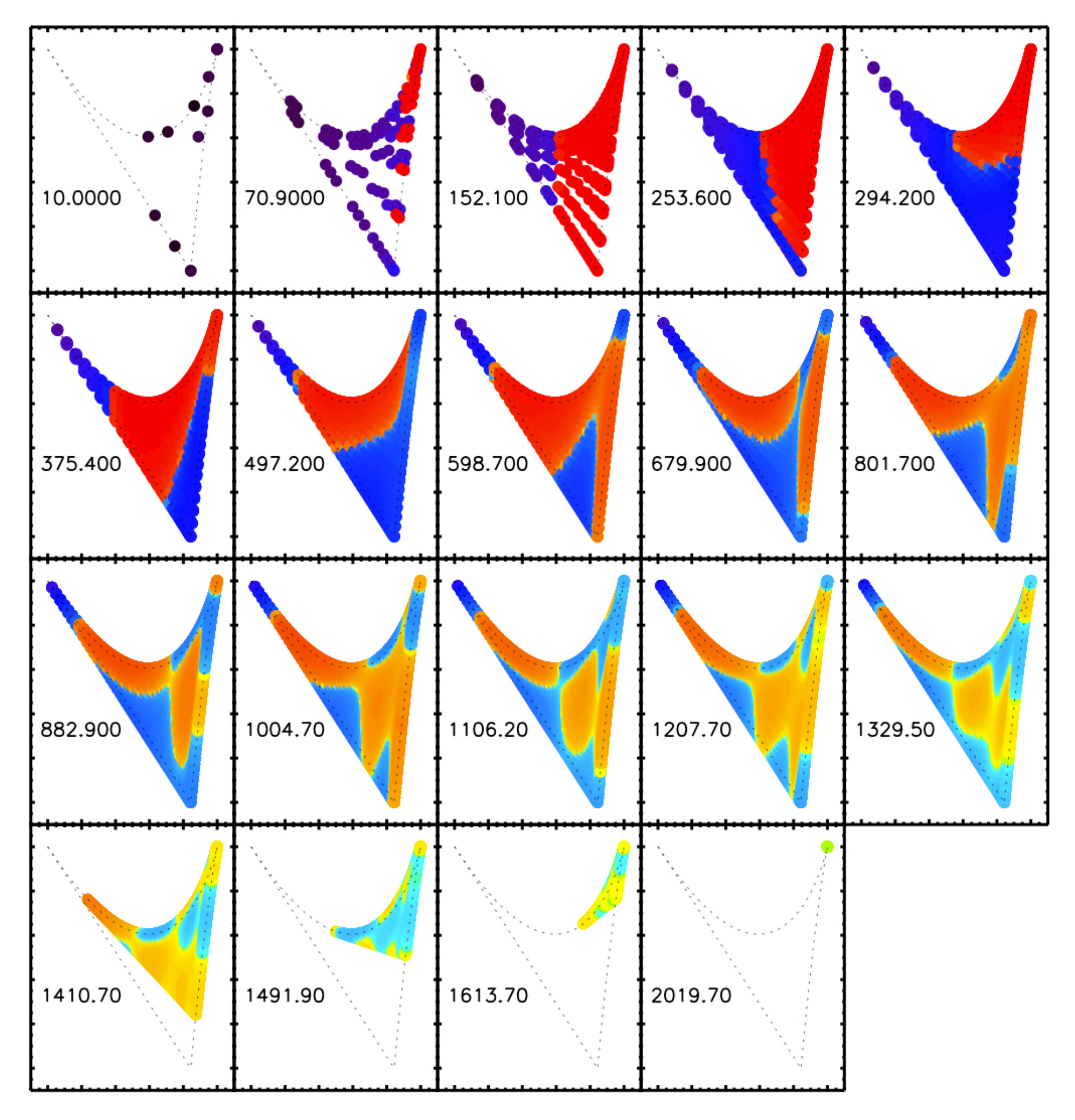}
\caption{Theoretical CMB angular bispectrum produced by the local
  model in the $(P,F,S)$-parametrisation. With chosen representative
  perimeters and asinh color scale from deep purple/black (most
  negative) to red/dark grey (positive). $P/3$ is given on each
  panel. The axes are the same as those in Fig. \ref{triparam}}
\label{fig:cmbparam}
\end{figure}

To illustrate the use of our $(P,F,S)$-parametrisation, we plot in
Fig. \ref{fig:cmbparam}\footnote{For space reasons we only include
  some of the bins in the figures in the paper. Full resolution plots
  with all perimeter bins are available upon request.} the theoretical
CMB bispectrum produced by the local NG model $f_\mathrm{NL}$,
computed through Eq. \ref{eq:cmbispth}.  The triangle perimeters, $P$,
vary between $P_\mathrm{min}=30$ (equilateral configuration with
$\ell_\mathrm{min}=10$) and $P_\mathrm{max}=6120$ (equilateral
configuration with $\ell_\mathrm{max}=2040$). We plot representative
perimeters tracing the building up of the bispectrum with scale,
giving $P/3$ on each panel. The color code scales from deep
purple/black (most negative) to red/dark grey (positive).

In the first panels for the smallest perimeters, few triangles are
present. As the perimeter increases the $(F,S)$ space is populated
starting with equilateral configurations (upper right corner) first to
reach squeezed configuration (upper left corner) later. Conversely in
the last panels for the largest perimeters, the resolution limit
$\ell_\mathrm{max}=2048$ limits the possible configurations, leaving
only the equilateral triangles in the end.  
\\The Sachs-Wolfe shape
(see Eq. \ref{eq:cmbispsw}) is visible at low perimeters, with the
colors (value of the bispectrum) varying horizontally with $S$ but not
vertically with $F$. We note that the strong negative values (deep
purple/black) are located in the near-squeezed configuration (upper
left corner). The sign of the radiation transfer function can be
traced via the equilateral triangles which are positive for $P/3 \sim
200$ (first acoustic peak) and become negative for $P/3 \sim 500$
(second acoustic peak) etc. The squeezed configuration in turn is
always negative as the smallest multipole has negative transfer
function via SW. The folded configuration has relatively large
negative values when the two smallest multipoles are in the first
acoustic peak while the biggest multipole is in the second acoustic
peak ($P\sim 900-1200 \Rightarrow P/3\sim300-400$). For larger
perimeters the structure becomes complex as several acoustic peaks
intervene.


\section{Physical prescription for the point-source
  angular bispectrum} \label{sect:prescription} 

%
%

In this section we develop a prescription which permits us to predict
the (bi)spectrum of point sources, starting from the simplest case of
a single randomly-distributed population to the case of multiple
clustered populations. This is a situation that is encountered in
current and future CMB analyses. Indeed until recently, CMB
experiments have focussed on frequencies where unclustered radio
sources are the only dominant kind of point sources, but the CMB is
also non-negligible at higher frequencies where an independent
population of dusty galaxies becomes important together with the SZ
signal of clusters. This is of particular relevance for Planck which
has a large frequency range covering both populations.
%
%
\subsection{Single source population: shot-noise contribution}
\label{sect:shot}
A source with flux $S$ enclosed in a pixel with solid angle
$\Omega_{\mathrm{pix}}$ yields a temperature variation $\Delta T =
k_\nu \frac{S}{\Omega_{\mathrm{pix}}}$, where $k_\nu =
\left.\frac{\partial B(\nu,T)}{\partial T}\right|_{T_\mathrm{CMB}}$,
$B(\nu,T)$ is the black-body spectrum and $T_\mathrm{CMB}$ is the CMB
mean temperature.

As shown in Appendix \ref{appendix:shot} the power spectrum of a source
population is given by:
\begin{equation}
C_\ell  =  C_\ell^\mathrm{clust} + C_\ell^{\mathrm{shot}} \,.
\label{eq:PS_cl}
\end{equation}
The discreteness of the sources produces a constant-term spectrum
$C_\ell^{\mathrm{shot}}$ which is usually named `Poissonian' because
the number of unclustered point sources is driven a priori by Poisson
statistics \citep{Sehgal2010}. The shot-noise term reads:
\begin{equation}\label{shotspec}
C_\ell^{\mathrm{shot}}  =  \frac{k_\nu^2}{4\pi} \times \!\int_0^{S_\mathrm{cut}} S^2
\,\frac{\mathrm{d}n}{\mathrm{d}S} \, \mathrm{d}S 
\end{equation}
where $\frac{\mathrm{d}n}{\mathrm{d}S}$ is the differential number counts of
sources and $S_\mathrm{cut}$ is the detection limit, i.e sources with $S>S_\mathrm{cut}$ are
detected and masked, the rest being unresolved.

The discreteness property of sources, when computing the three-point
correlation function, yields a statistically isotropic
angular bispectrum constant with $\ell$:  
\begin{equation}\label{shotbisp}
b^{\mathrm{shot}}_{\ell_1\ell_2\ell_3}  =  \frac{k_\nu^3}{4\pi} \times
\int_0^{S_\mathrm{cut}} S^3 \,\frac{\mathrm{d}n}{\mathrm{d}S} \, \mathrm{d}S 
\end{equation}
for $\ell_i \ne 0$. Equations (\ref{eq:PS_cl}), (\ref{shotspec}) and
(\ref{shotbisp}) are derived in more detail in Appendix
\ref{appendix:shot}.

The case of sources randomly and independently distributed on the sky
is that of the radio sources at CMB frequencies. The correlation
vanishes and the total (bi)spectrum is equal to the shot-noise
(bi)spectrum. The distribution of the sources is that of a white-noise
entirely characterised by the one-point probability distribution. In
this case, Eqs.  (\ref{shotspec}) and (\ref{shotbisp}) for the shot noise
contribution can be reformulated simply in terms of temperature
variations where the white-noise spectrum and bispectrum are related
to the variance $\sigma^2$ and skewness $\kappa_3$ of $\Delta T$:
\begin{equation}\label{specshotT}
C^{\mathrm{white}}_\ell  =  \sigma^2 \, \Omega_{\mathrm{pix}}
\end{equation}
\begin{equation}\label{bispshotT}
b^{\mathrm{white}}_{\ell_1 \ell_2 \ell_3}  =  \kappa_3 \, \Omega^2_{\mathrm{pix}}
\end{equation}
with $\sigma^2= \langle (\Delta T-\langle \Delta T\rangle)^2\rangle$ and
$\kappa_3= \langle (\Delta T-\langle\Delta T\rangle)^3\rangle$.
%
%
\subsection{Single source population with correlations}\label{sect:onepop}
The effect of clustering of a single population of sources, namely
radio sources, on the bispectrum was pioneered by \cite{Argueso2003} who
proposed a prescription to address this issue.  The elements entering
the prescription are the number counts of sources and a theoretical or
observational correlation function w($\theta$). Defining
\begin{equation}
P(k)_{\mathrm{clust}} = 2\pi \int w(\theta) J_0(k\theta) \mathrm{d}\theta
\end{equation}
where $\theta$ is the distance on the flat patch and $J_0$ the Bessel
function of the first kind and of order zero, the prescription is: 
\begin{equation} \label{argueso}
\delta_{\vec{k}}(\mathrm{tot})=\sqrt{\frac{P(k)_{\mathrm{clust}}+P(k)_{\mathrm{shot}}}{P(k)_{\mathrm{white}}}}\times\delta_{\vec{k}}(\mathrm{white}) 
\end{equation}
where $\delta_{\vec{k}}$ are the Fourier coefficient of the map,
`shot' and `white' refer respectively to the shot-noise and
white-noise process. Then Argueso et al. compute the angular bispectrum from
simulated square maps based on their prescription.

In our study, we have extended the above-described prescription to
analytically derive the full-sky angular bispectrum. The full-sky scale-maps
read: 
\begin{equation*}
T_{\ell}^{\mathrm{tot}}(\mathbf{n}) =
\sqrt{\frac{C_{\ell}^{\mathrm{clust}}+C_\ell^{\mathrm{shot}}}{C_\ell^{\mathrm{white}}}}
\times T_{\ell}^{\mathrm{white}}(\mathbf{n}) 
\end{equation*}
which reduces to Eq. (\ref{argueso}) in the flat-sky approximation
$\ell=2\pi k$, and assuming $P(k)$ does not vary much within a $k$
bin. 
\\So the power spectrum is given by:
\begin{equation*}
C_{\ell}^{\mathrm{tot}} = C_{\ell}^{\mathrm{clust}}+C_{\ell}^{\mathrm{shot}} .
\end{equation*}
We remind the reader that the Argueso et al.'s prescription is
specifically made to meet the above relation, and that
$C_{\ell}^{\mathrm{shot}}=C_{\ell}^{\mathrm{white}}=\mathrm{const}$. 

More interestingly the bispectrum is found to be:
\begin{eqnarray}\label{bispargueso}
\nonumber b_{\ell_1 \ell_2 \ell_3}^\mathrm{tot} & = &
\sqrt{1+\frac{C_{\ell_1}^{\mathrm{clust}}}{C_{\ell_1}^{\mathrm{white}}}}
\sqrt{1+\frac{C_{\ell_2}^{\mathrm{clust}}}{C_{\ell_2}^{\mathrm{white}}}} 
\sqrt{1+\frac{C_{\ell_3}^{\mathrm{clust}}}{C_{\ell_3}^{\mathrm{white}}}} \\
& \ & \qquad\qquad\qquad\qquad\qquad\qquad \times \ 
b_{\ell_1 \ell_2 \ell_3}^{\mathrm{white}} \\ 
\nonumber & = \alpha & \sqrt{C_{\ell_1}^{\mathrm{tot}} \,
  C_{\ell_2}^{\mathrm{tot}} \, C_{\ell_3}^{\mathrm{tot}}} 
\end{eqnarray}
with $b_{\ell_1 \ell_2 \ell_3}^{\mathrm{white}}=\mathrm{const}$ and
$\alpha = \frac{b_{\ell_1 \ell_2
    \ell_3}^{\mathrm{white}}}{\sqrt{C_{\ell_1}^{\mathrm{white}}C_{\ell_2}^{\mathrm{white}}C_{\ell_3}^{\mathrm{white}}}}$.  

Equation (\ref{bispargueso}) of the bispectrum relates to the
clustered power spectrum, or conversely the correlation function
entering in the prescription. 
%
%
\subsection{Two populations of sources with clustering}\label{sect:twpop}
The previous prescription, Eq. (\ref{argueso}), describes well a single
point source population. However, it fails at describing the cases
where more than one population contribute to the signal. In
particular, the case of the present generation of CMB experiments
observing the CMB from low (30 GHz) to high (860 GHz) frequencies calls
for an appropriate prescription which deals with independent source
populations. Indeed, early results from the Planck mission
\citep{Planck-Collaboration-statradio}
\citep{Planck-Collaboration-radioSED} show that radio and IR galaxies
both contribute at frequencies of 100 GHz and above. In the following
we thus extend and generalise the prescription accordingly.
 
We hence define the harmonic coefficients as:
\begin{equation} \label{lacasa}
a_{\ell m}^{(\mathrm{tot})} = a_{\ell m}^{(\mathrm{white, 1})}
+
\sqrt{\frac{C_\ell^{\mathrm{clust}}+C_\ell^{\mathrm{shot}}}{C_\ell^{\mathrm{white,2}}}} 
\times a_{\ell m}^{(\mathrm{white, 2})}
\end{equation}
where $a_{\ell m}^{(\mathrm{white, 1})}$ and $a_{\ell m}^{(\mathrm{white,  2})}$ 
are independent realisations of white-noise with different
number counts. Index 1 refers to the radio population and 2 to the
infrared population. The spectrum has then a form similar to the case
of a single source 
population: 
\begin{equation*}
C_{\ell} =  C_{\ell}^{\mathrm{clust}} + C_{\ell}^{\mathrm{shot,1+2}}
\end{equation*}
But the angular bispectrum differs, and reads:
\begin{equation}\label{bisplac}
b^\mathrm{tot}_{\ell_1 \ell_2 \ell_3} = b^{\mathrm{white, 1}}_{\ell_1 \ell_2 \ell_3} +
\sqrt{\frac{C_{\ell_1}^{\mathrm{tot,2}}}{C_{\ell_1}^{\mathrm{white,2}}}
\,\frac{C_{\ell_2}^{\mathrm{tot,2}}}{C_{\ell_2}^{\mathrm{white,2}}}
\,\frac{C_{\ell_3}^{\mathrm{tot,2}}}{C_{\ell_3}^{\mathrm{white,2}}}}
\times b^{\mathrm{white, 2}}_{\ell_1 \ell_2 \ell_3} 
\end{equation}

For illustration, let us compare equations (\ref{bispargueso}) and
(\ref{bisplac}) in the equilateral configuration with white-noises of
both populations derived from the same number counts, and neglecting
the shot-noise of the second population: 
\begin{equation}\label{equiargueso}
b^\mathrm{1pop}_{\ell \ell \ell} = \left(1 +
\frac{C_{\ell}^{\mathrm{clust}}}{C_{\ell}^{\mathrm{white}}}\right)^{3/2}
\; b^\mathrm{white}_{\ell \ell \ell} 
\end{equation}
\begin{equation}\label{equilacasa}
b^\mathrm{2pop}_{\ell \ell \ell} = \left(1 +
\left(\frac{C_{\ell}^{\mathrm{clust}}}{C_{\ell}^{\mathrm{white}}}\right)^{3/2}\right) 
\; b^\mathrm{white}_{\ell \ell \ell} 
\end{equation}
The difference between these two formula is maximal when
$C_{\ell}^{\mathrm{clust}}/C_{\ell}^{\mathrm{white}} \simeq 1$ and can
be up to 40\%, illustrating the need to properly account for the
different populations. 

The two-population case, representative of the CMB context in the
frequency range of interest, can be straightforwardly generalised to
more point-source populations with or without clustering properties.

%
%
\section{Results}\label{sect:resultsehgal}

In this section we present the bispectra of radio and IR sources
computed on simulations described below, showing the configuration
dependence of the angular bispectrum and its frequency behaviour.

\subsection{Data used}

For our analysis, we used the all-sky simulated maps\footnote{The
  frequency maps are available on Lambda website,
  http://lambda.gsfc.nasa.gov/toolbox/tb\_cmbsim\_ov.cfm} of the IR
and radio point sources provided by \cite{Sehgal2010} at 30, 90, 148,
219 and 350 GHz. We provide here a brief summary of the map
construction. For a detailed description, we refer the reader to the
above-cited article. The maps are based on N-body simulations of the
large scale structure, with a volume 1000 $\mathrm{h}^{-1} $Mpc
on a side, produced using a tree-particle mesh code. Dark matter (DM)
haloes are identified and are then populated with infrared and radio
galaxies. The model for the radio sources is adapted so that the radio
luminosity function matches that of the observed radio sources at 151
MHz.

The IR-source model was partially based on \cite{Righi2008}.  The DM
haloes are populated with galaxies of given luminosities taking into
account a Poisson term and a correlation term. The model was
constructed so that it is compatible with constraints on the
luminosity function, the source counts and the fluctuations from
SCUBA, BLAST, Spitzer and ACBAR (see \cite{Sehgal2010} for
details). However, the simulation of IR sources, used here, does not
account for the most recent observational constraints from ACT, SPT
Herschel and Planck results.

Maps of the different astrophysical contributions, in HEALPix format
at Nside=8192, were produced by replication of one simulated octant of
the sky. This procedure does not properly account for the signal at
the largest scales especially up to the octopole, $\ell=3$ ; it also
introduces excess power at $\ell \le 300$ for infrared sources as
discussed by \cite{Sehgal2010}, but the power is correct for higher
multipoles. The produced maps have half-arcminute resolution but we
degraded them to Nside=1024 and used a uniform binning
$\Delta\ell=64$, to speed up computations. We checked that this
procedure does not introduce a bias by comparing the binned spectrum
in the degraded map to the unbinned spectrum in the original map and
find excellent agreement. The octant replication in the map-making
translates mainly into a lack of power in the first bin (centered
around $\ell=32$) which is hence discarded in later plots.

%
%
\subsection{Radio source characterisation}\label{sect:radresults}
We first investigate the bispectrum dependence on the configurations
at a given frequency. We plot the angular bispectrum in four commonly used
configurations, namely equilateral $(\ell,\ell,\ell)$, isosceles
orthogonal $(\ell,\ell,\sqrt{2}\ell)$, isosceles flat
$(\ell,\ell,2\ell)$, and squeezed $(\ell_\mathrm{min},\ell,\ell)$
configurations. They furthermore sample rather well the configuration
space (see Fig. \ref{triparam}).

\begin{figure}
\centering
\includegraphics[width=8.5cm]{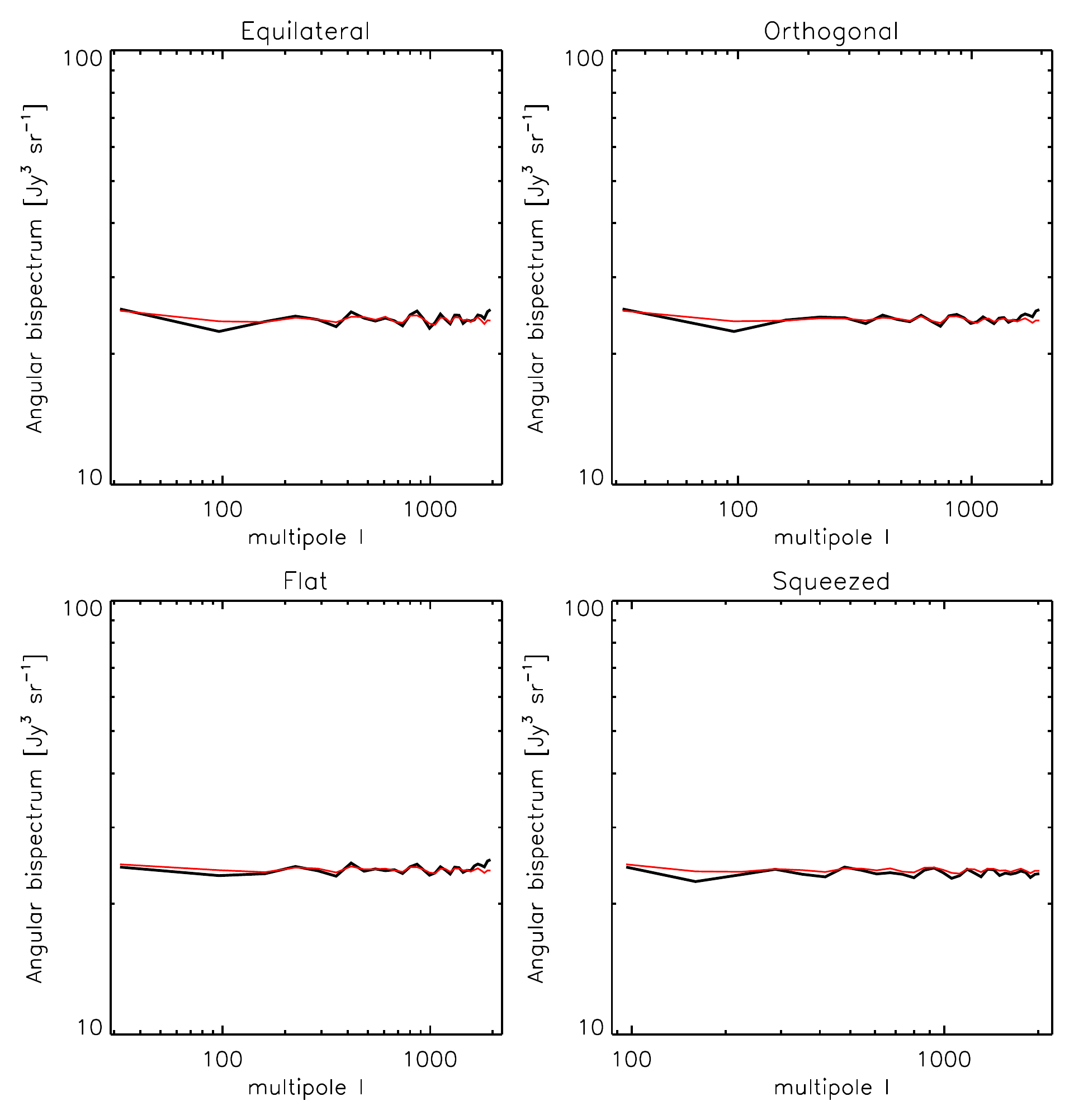}
\caption{Thick black line: Radio angular bispectrum at 90 GHz in different
  configurations: equilateral $(\ell,\ell,\ell)$, isocele orthogonal
  $(\ell,\ell,\sqrt{2}\ell)$, isocele flat $(\ell,\ell,2\ell)$, and
  squeezed $(\ell_\mathrm{min},\ell,\ell)$. The thin red line is the fit with
  prescription. The unit of the bispectra is Jy$^3\cdot \mathrm{sr}^{-1}$}
\label{speconfrad090}
\end{figure}

We see in Fig. \ref{speconfrad090}, black thick line, that the
bispectrum is constant. This result is independent of the
frequency. Moreover the value of the constant is independent of the
configuration. This is what we expect from white noise and it shows
that radio sources are indeed randomly distributed over the sky.

\begin{figure}
\centering
\includegraphics[width=8.5cm]{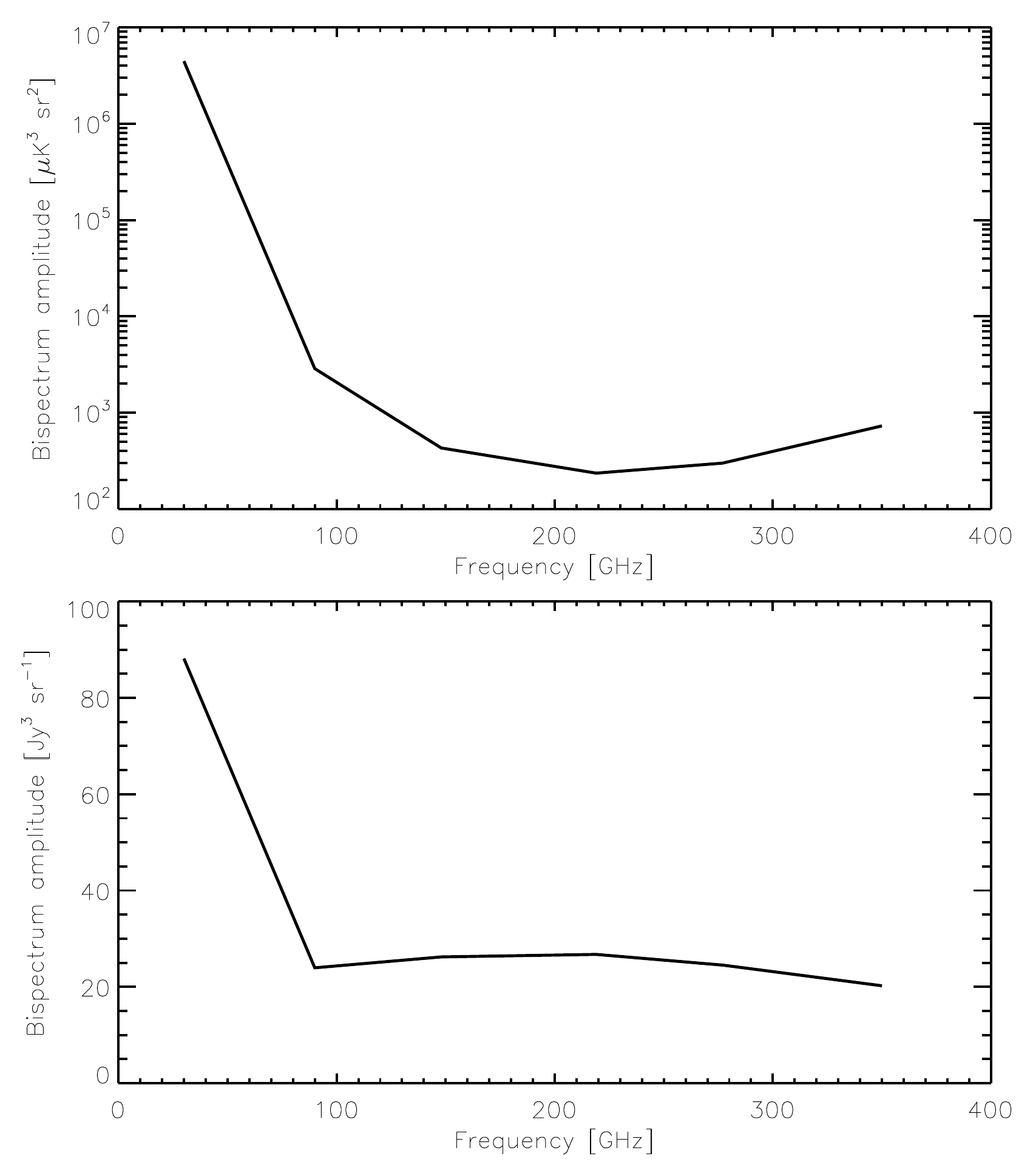}
\caption{Amplitude of the radio angular bispectrum, averaged over all
  configurations, at 30, 219, 148, 277, 90, and 350 GHz, in
  temperature units for the upper panel and flux units for the bottom
  panel.}
\label{fig:amplRAD}
\end{figure}

We show in Fig. \ref{fig:amplRAD} the dependence with frequency of the
bispectrum amplitude averaged over all the configurations, in
equivalent temperature unit for the upper panel and flux unit for the
bottom panel. The amplitude is maximal at the lowest frequency 30 GHz,
it then decreases to become mostly flat above 90 GHz because of
free-free emission and inverted spectra sources.

\begin{figure}
\centering
\includegraphics[width=8.5cm]{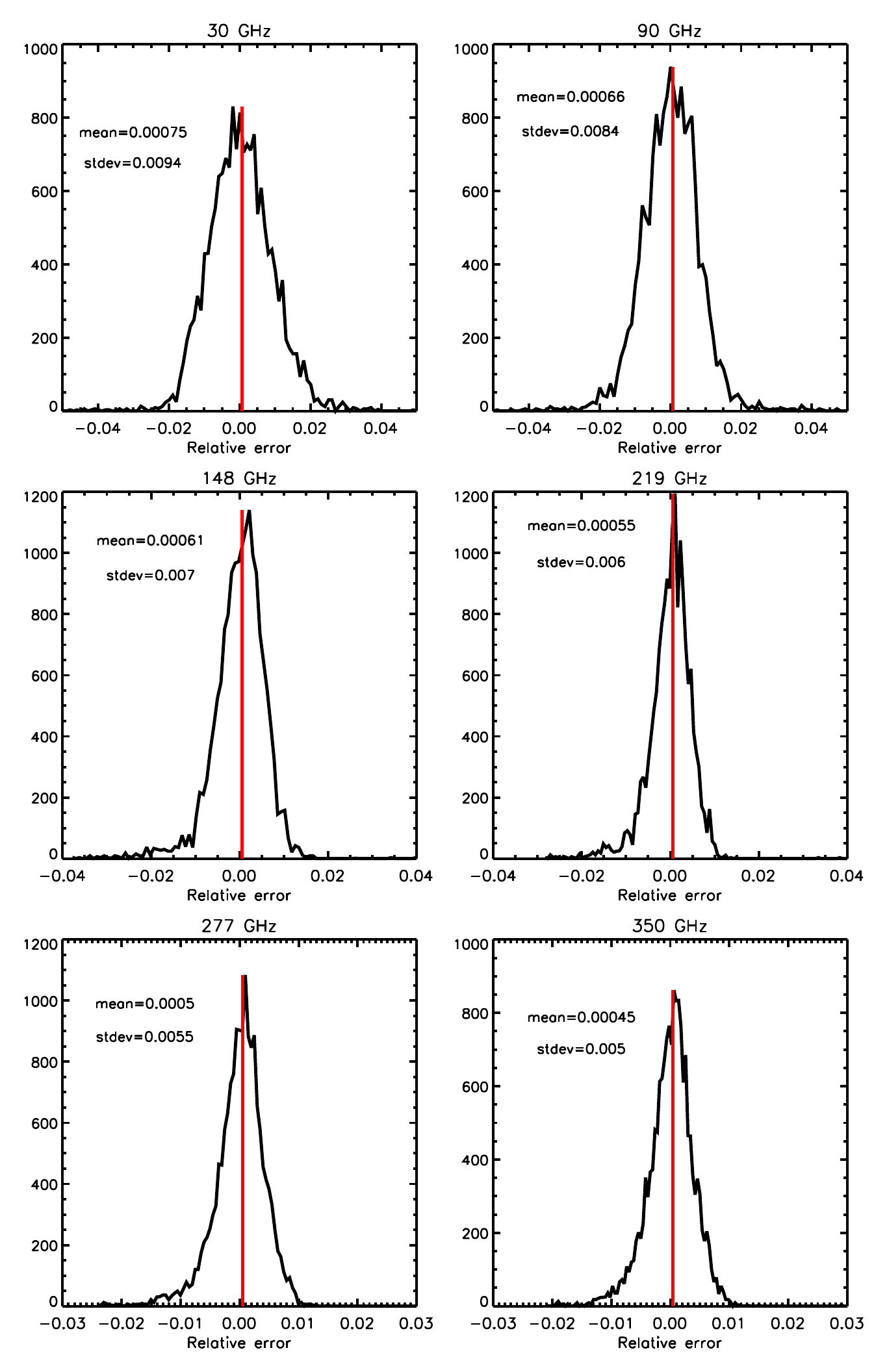}
\caption{Relative error distribution with the prescription for radio
  sources alone. The vertical bar shows the mean of the distribution} 
\label{histerrorfitrad}
\end{figure}

The bispectrum was fitted with the one population prescription
described in Sect. \ref{sect:onepop}, using the power spectrum
extracted from the simulations and the multiplicative constant which
minimizes the $\chi^2$ of the observed bispectrum to the prescribed
one with error bars from Wick's expansion (see Appendix
\ref{appendix:wngvar}).  The fit of the bispectrum with the
prescription is very good: Fig. \ref{histerrorfitrad} shows that the
relative error (exact to fit) lies in the range between -2\% and +2\%
with a mean relative error always less than 1$\tcperthousand$.


\subsection{IR source characterisation}\label{sect:irresults}
\begin{figure}
\centering
\includegraphics[width=8.5cm]{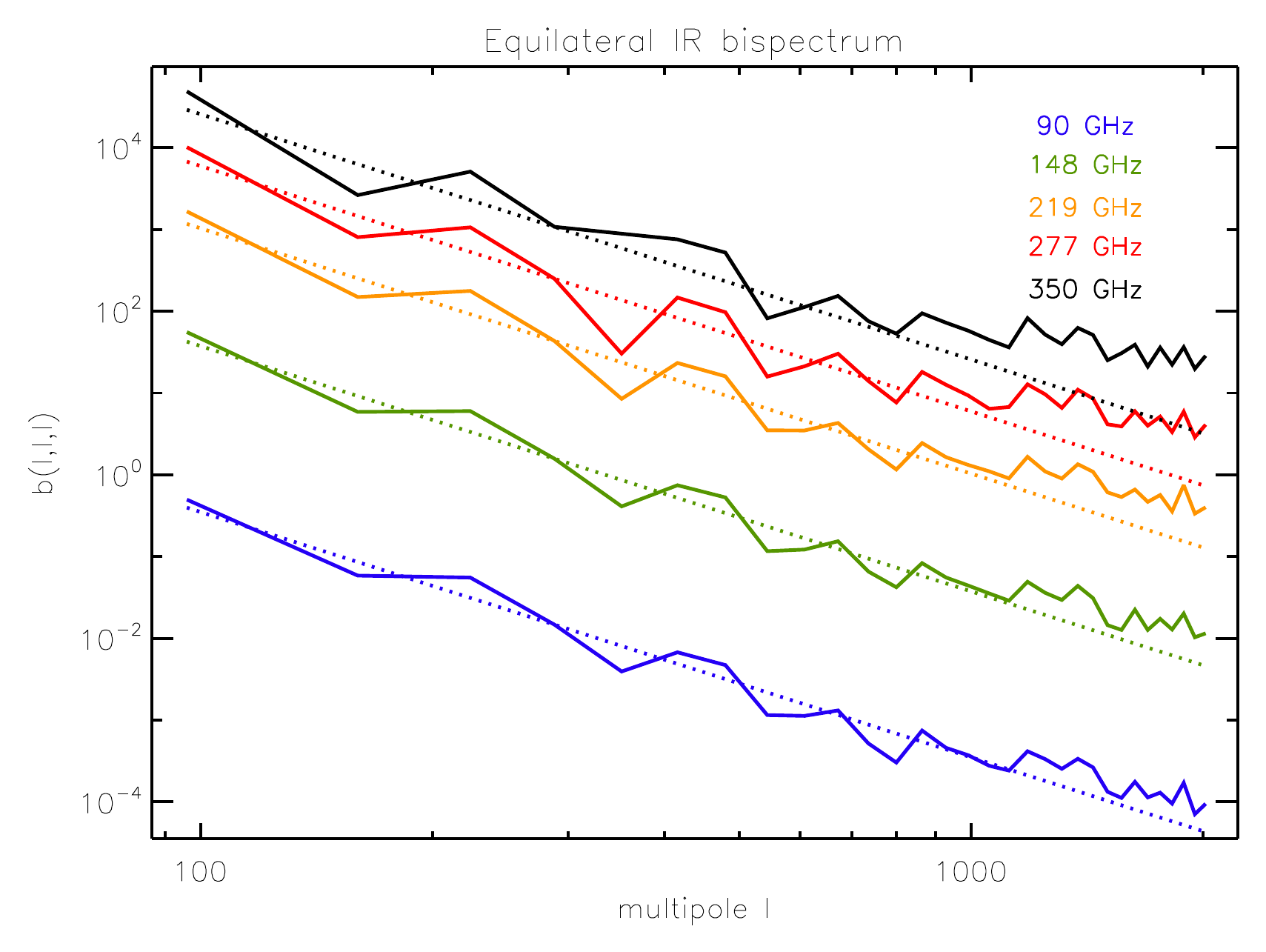}
\caption{Infrared equilateral bispectrum at 90, 148, 219, 277, and 350
  GHz from bottom to top. Dotted corresponding lines are the power-law
  fit}
\label{equibisp-IR}
\end{figure}

Figure \ref{equibisp-IR} shows the amplitude of the angular bispectrum in the
equilateral configuration for the different frequencies. Dusty galaxy
emission peaks at IR frequencies, and plummets in the radio domain and
so does the amplitude of the bispectrum. The bispectrum decreases with
$\ell$, well fitted by a power-law for $100\leq \ell \leq 1000$, and
with a flattening at higher multipoles. The decrease is expected from
the clustering of IR galaxies on large scales. We found that the
slopes were about the same for equilateral, flat and orthogonal
configurations except for the squeezed triangles. Indeed, for the latter
$\ell_1$ is fixed, while for the other configurations
$\ell_1,\ell_2,\ell_3$ are all proportional to $\ell$.

The flattening of the bispectrum at high multipoles, indicative of the
shot-noise contribution, occurs at lower multipoles with increasing
frequency (e.g. at 350 GHz the bispectrum deviates from a power law at
$\ell \sim 1000$ while at 90 GHz it is not before $\ell \sim
1500$). This is explained by the contribution of the high-flux
galaxies, in Sehgal et al.'s simulations, that accounts for the
shot-noise ($b^\mathrm{shot} \sim \int S^3
\frac{\mathrm{d}n}{\mathrm{d}S} \mathrm{d}S$) and at the same time
have a steeper emission than the galaxies accounting for the
clustering term.

\begin{figure}
\centering
\includegraphics[width=8.5cm]{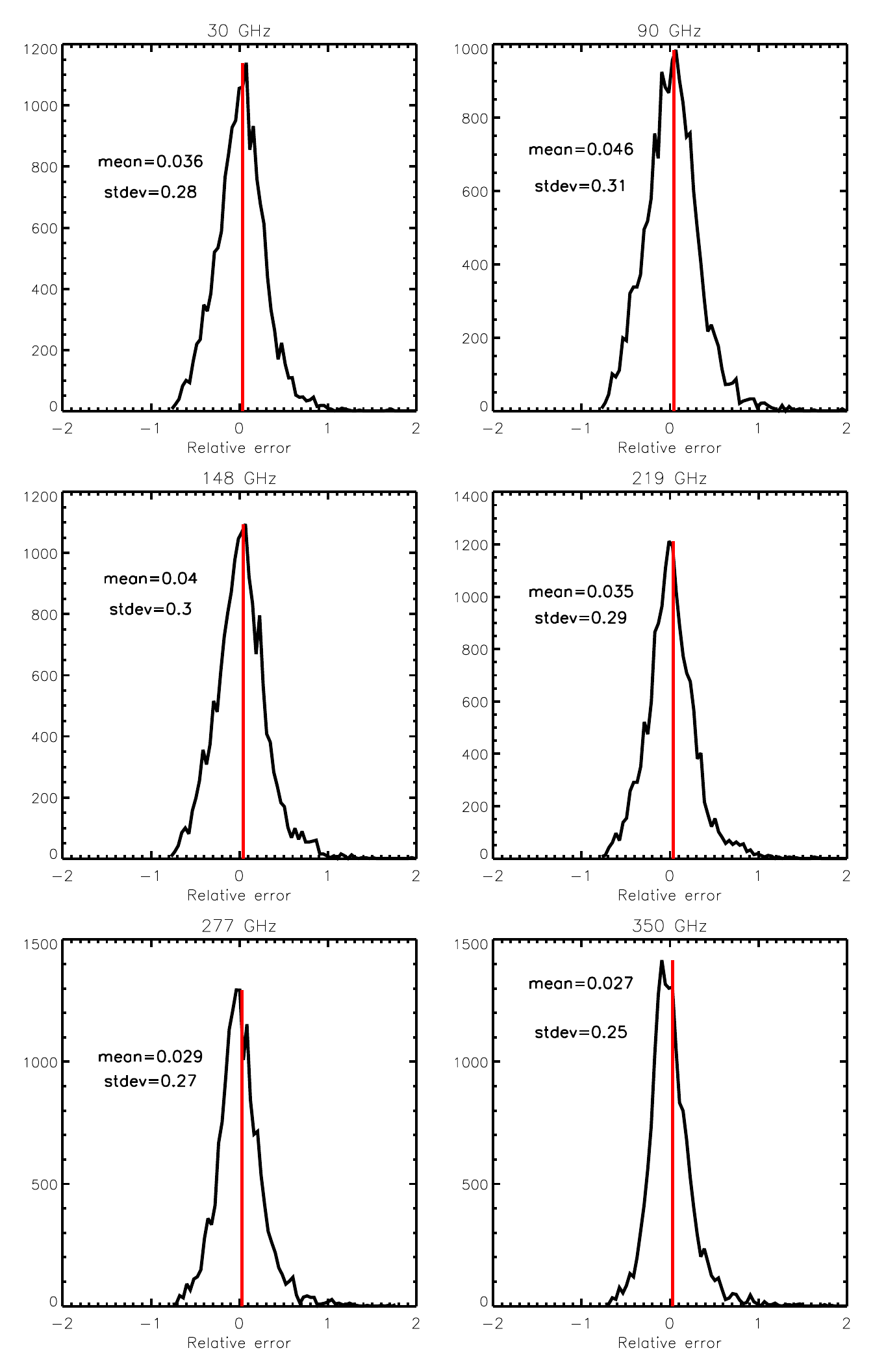}
\caption{Relative error distribution with the prescription for
  infrared sources alone. The vertical bar shows the mean of the
  distribution. The quoted standard deviation at 350 GHz is
    computed after discarding one unique negative bispectrum outlier
    at 400$\sigma$.}
\label{histerrorfitir}
\end{figure}

The bispectrum was fitted with the one population prescription
described in Sect. \ref{sect:onepop}, using the multiplicative
constant which minimizes the $\chi^2$, as previously for radio
sources. We show in Fig.  \ref{histerrorfitir} how the bispectrum of
the IR sources compares with the prescribed one. We see that the
bispectrum obtained with the prescription is good, with a mean
relative error always $\leq 5\%$ and a standard deviation $\leq
31\%$. At 350 GHz an outlier at -400$\sigma$ was discarded for the
computation of the standard deviation.  Figures \ref{histerrorfitrad}
and \ref{histerrorfitir} show the distribution of the relative error
between the bispectrum derived from the prescription and the actual
bispectrum measured in the simulated maps. In other words, it exhibits
departures from the predicted bispectrum values. The dispersion of
these relative errors is larger for the IR sources
(Fig. \ref{histerrorfitir}) than for the radio sources
(Fig. \ref{histerrorfitrad}) at all the frequencies. This behaviour is
not an indication of a mismatch between the predicted and the actual
bispectra; it relates to the intrinsic sample variance of both the IR
and radio bispectra. The IR sources being weakly non-Gaussian (the
value for $\alpha$ defined in Eq. (\ref{bispargueso}) is $3\times
10^{-3}$ for the IR sources, compared to 0.3 for radio sources) the
variance of their bispectrum is indeed large compared to the
bispectrum value (see Appendix \ref{appendix:wngvar}).

\subsection{Total contribution from IR and radio source
  populations}\label{sect:iradresults} 

\begin{figure}
\centering
\includegraphics[width=8.5cm]{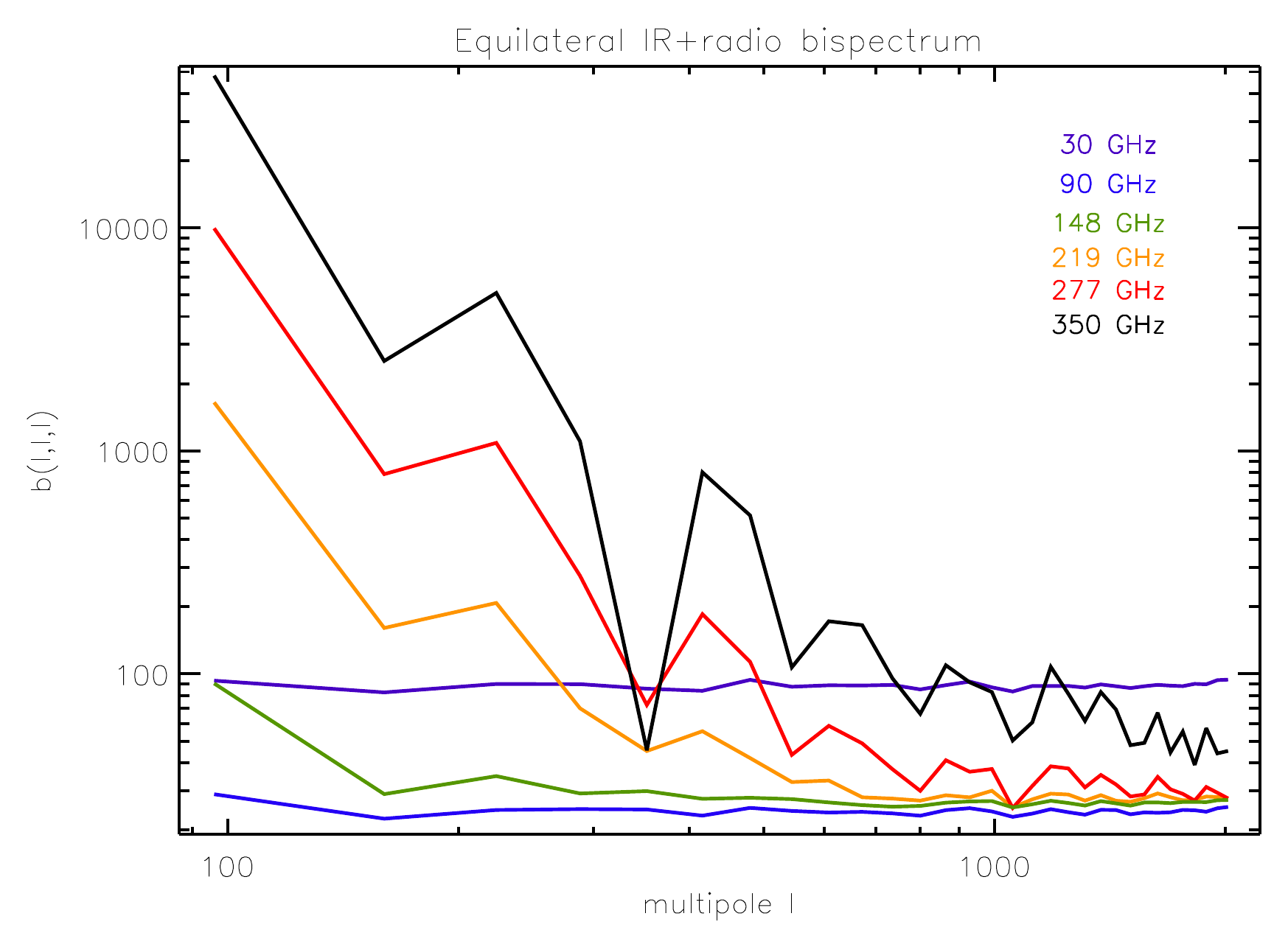}
\caption{Infrared + radio orthogonal bispectrum at 30, 90, 148, 219,
  277, and 350 GHz. The purple flat upper line is 30 GHz, the blue flat lowest
  line is 90 GHz, more variable decreasing lines from bottom to top
  are 148, 219, 277, and 350 GHz} 
\label{equibisp-IRAD}
\end{figure}

We now present the results when the two populations of sources
contribute to the signal at the frequencies 30, 90, 148, 219, 277, and
350 GHz. To do so, we simply add the simulated maps at each
frequency. \\ We illustrate the angular bispectrum dependence on frequency for
one single configuration, namely equilateral, see Fig.
\ref{equibisp-IRAD}.

\begin{figure}
\centering
\includegraphics[width=8.5cm]{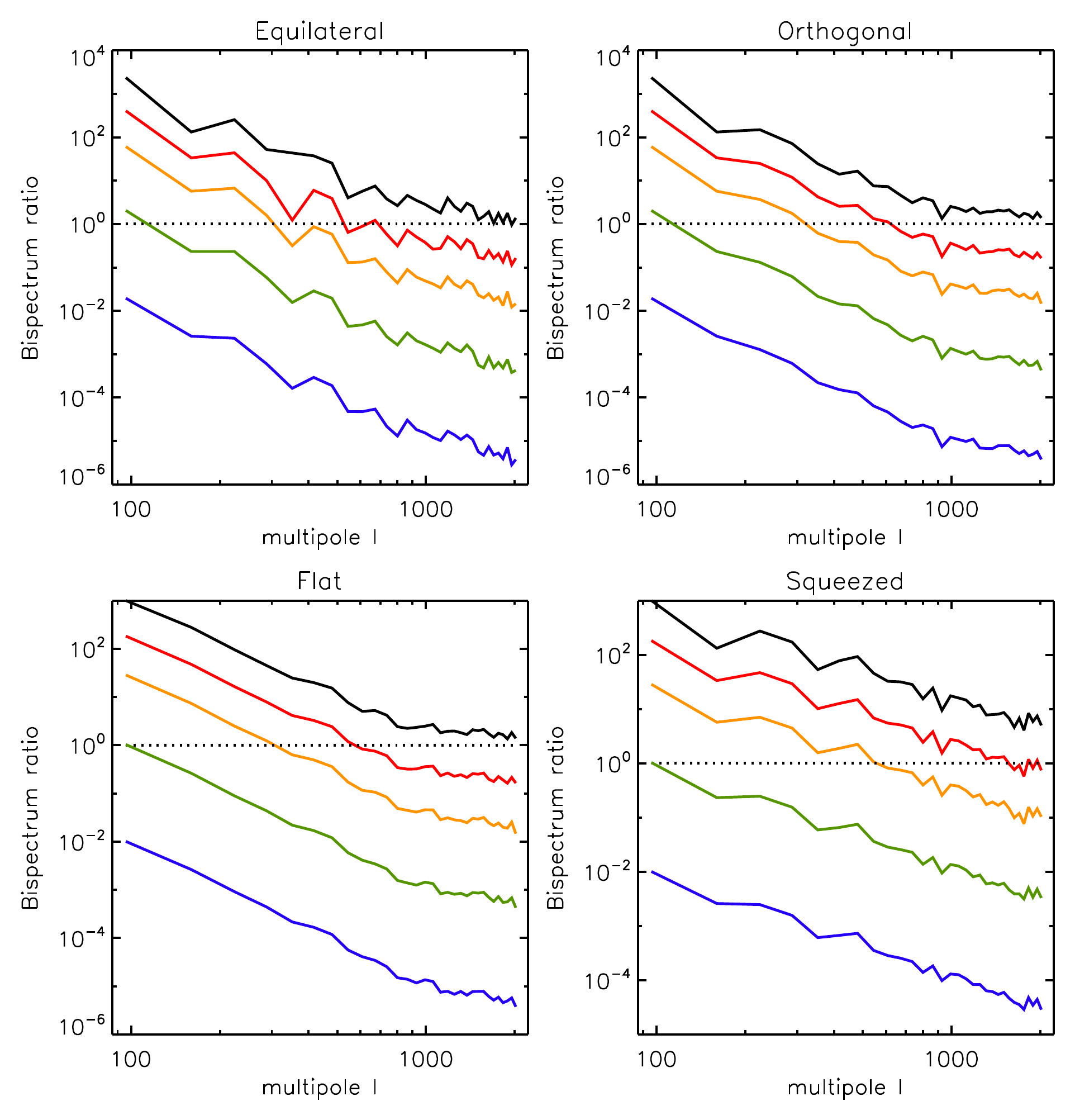}
\caption{Ratio of the IR angular bispectrum to the radio one at 90, 148, 219,
  277, and 350 GHz from bottom to top. The dotted line indicates equality.} 
\label{speconfratio}
\end{figure}

The frequency behaviour is as expected from an independent combination
of the IR and RAD bispectra. The radio source contribution dominates
at low frequencies 30 and 90 GHz (blue and purple lines) and its
bispectrum is flat. Infrared galaxies dominate at the highest
frequencies 277 and 350 GHz (black and red upper lines) and show the
characteristic power-law dependence due to clustering followed by a
flattening of the bispectrum. At intermediate frequencies both
populations contribute to the signal. The clustering-induced term of
IR-galaxies dominates on large angular scale while the random-noise
term of radio-galaxies dominates at small angular scale. The
cross-over between radio and IR-galaxy bispectra is shifted to higher
$\ell$s with increasing frequency. It is worth noting in
Fig. \ref{speconfratio} that at the lowest multipoles and at highest
frequencies, the IR galaxies produce a bispectrum at least 10 times
more important than the radio sources.

\begin{figure}
\centering
\includegraphics[width=8.5cm]{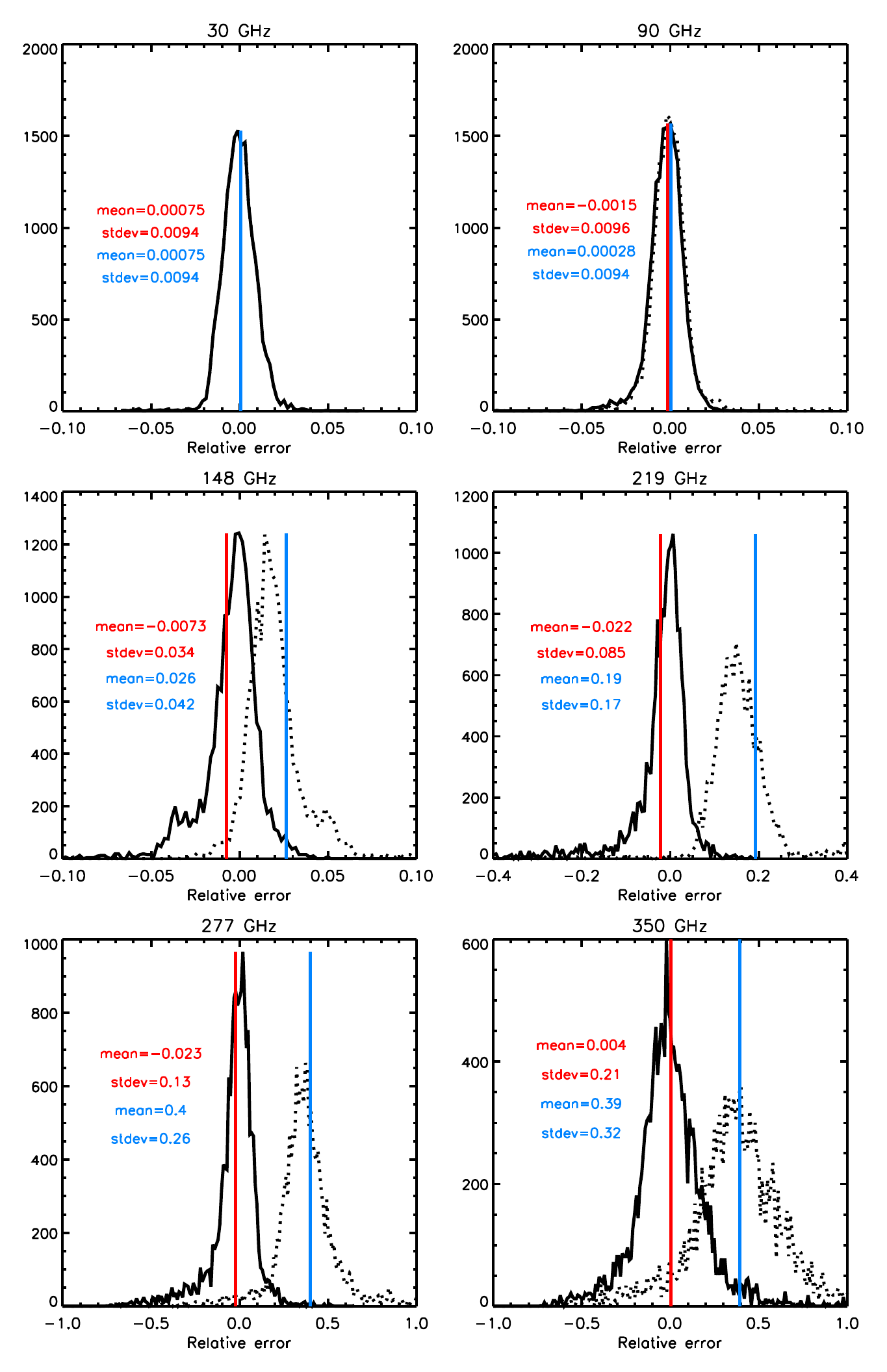}
\caption{Relative error distribution with the prescriptions for the
  combination of infrared and radio sources. The dotted line is
  obtained with the single population prescription, the solid line
  with the two populations prescription. Vertical bars show the mean
  of the distribution, blue for the single population prescription and
  red for the two populations prescription.}
\label{histerrorfitirad}
\end{figure}

We illustrate the angular bispectrum dependence on configurations at
350 GHz, see Fig. \ref{speconfirad350} thick black line. The errors
bars were computed in the weak NG approximation see Appendix
\ref{appendix:wngvar}. We note that infrared-radio cross-over occurs
at about the same scale for the equilateral, orthogonal and flat
configurations, but is at higher $\ell$ in squeezed
configuration. This is expected because the squeezed IR bispectrum
decreases more slowly than other configurations, one of the multipoles
being fixed. Figure \ref{speconfirad350} also displays (thin red line)
the bispectrum computed with two-population prescription derived in
Sect. \ref{sect:prescription}, i.e. adding up independently the
prescription for radio sources and infrared sources derived in the
previous sections.  This is compared to a bispectrum computation
considering only a single population (thin blue line). From
Fig. \ref{speconfirad350}, and Fig.  \ref{histerrorfitirad} showing
the distribution of relative errors with respect to these two
prescriptions, we see that the two-population prescription performs
much better than the single-population prescription. The former
captures well the overall shape of the bispectrum and it adjusts
particularly well the high $\ell$s. As a matter of fact, the mean
relative error is lower than 2.5\% up to 350 GHz and the dispersion
increases from 1\% at 30 GHz to 21\% at 350 GHz.

As expected, at 30 and 90 GHz the two prescriptions give same results
since radio sources totally dominate the signal. At higher
frequencies, both the mean errors and the dispersions derived using
the two-population prescription are smaller than those obtained with
the single-population prescription. Interestingly enough, at the
highest frequency (350 GHz) where IR emission from galaxies is
dominant, the single-population prescription is not satisfactory.  As
a matter of fact, configurations with at least one high multipole
dominate the distributions (e.g. 7/8th of the configurations have at
least one $\ell_i \geq 1000$). At 350 GHz these $\ell$s are dominated
by infrared shot-noise, so the computation of the prescriptions
combine the IR shot noise and radio spectrum which are both flat. The
radio emission is subdominant compared to infrared shot-noise but
nevertheless not negligible so the single-population prescription
leads to an overestimate of the total bispectrum. This is clearly
visible in Fig. \ref{speconfirad350} where the single population
prescription (blue thin line) is systematically higher than the
computed bispectrum (black thick line) and than the two-population
prescription (red thin line), particularly at high multipoles.

\begin{figure}
\centering
\includegraphics[width=8.5cm]{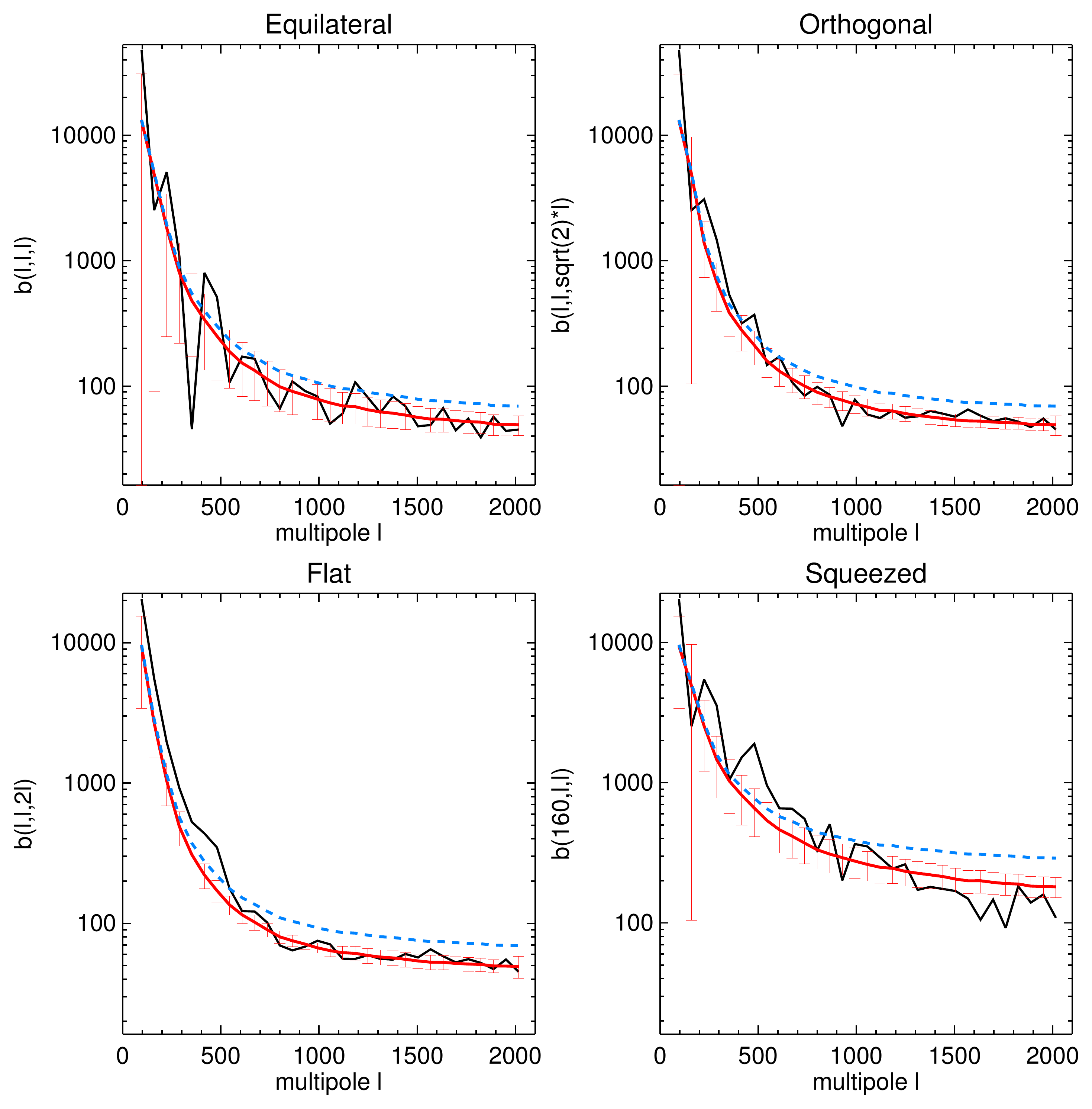}
\caption{Infrared + radio bispectrum at 350 GHz in different
  configurations, with error bars. The thick black line is the computed
  bispectrum. Smoother solid lines are the fit with prescriptions: the
  upper blue line with the single-source prescription, and the lower
  red line with the two-sources prescription.}
\label{speconfirad350}
\end{figure}

\section{Consequences on non-Gaussianity measures}\label{sect:ngcsqce}

\subsection{$(P,F,S)$ parametrisation}

\begin{figure}
\centering
\includegraphics[width=8.5cm]{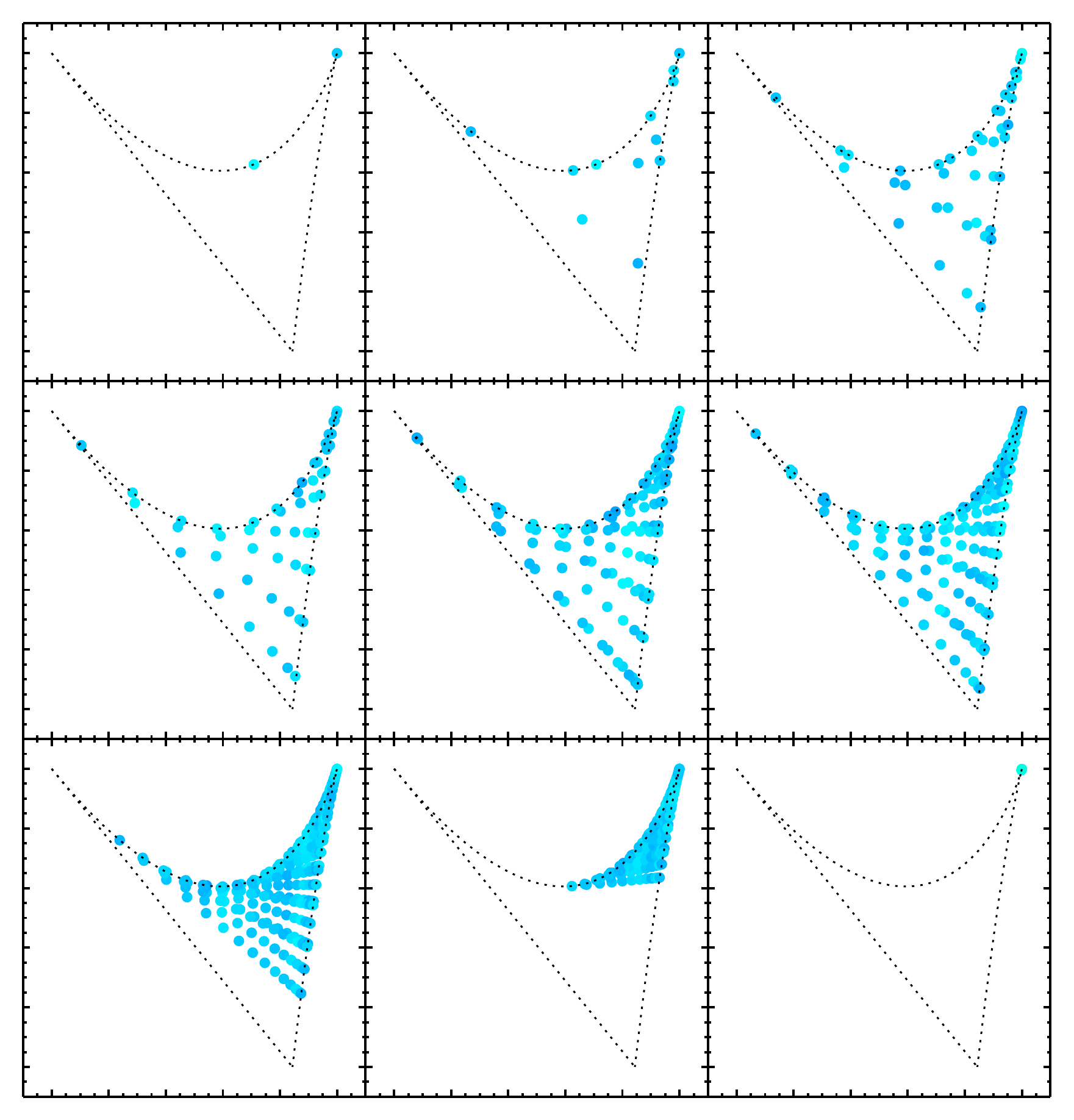}
\caption{Radio bispectrum in the $(P,F,S)$ parametrisation at 90
  GHz. Each plot is a slice of constant perimeter $P$, the value of the
  bispectrum is encoded in a logarithmic color scale from violet-blue
  to red. The axes are the same as those in Fig. \ref{triparam}}
\label{fig:paramrad}
\end{figure}

\begin{figure}
\centering
\includegraphics[width=8.5cm]{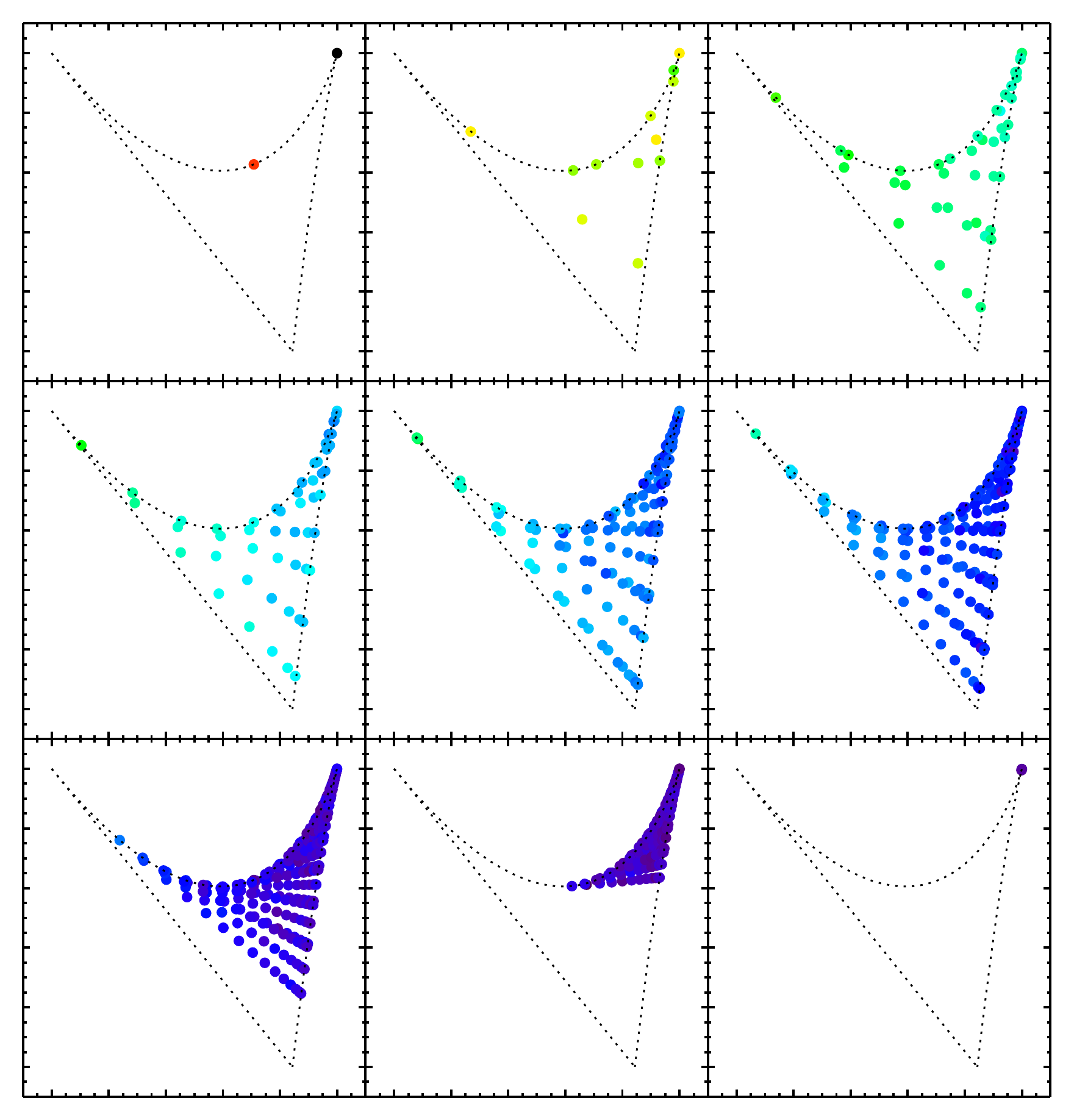}
\caption{IR bispectrum in the $(P,F,S)$ parametrisation at 148 GHz. The axes are the same as those in Fig. \ref{triparam}}
\label{fig:paramir}
\end{figure}

The parametrisation of the angular bispectrum proposed in
Sect. \ref{sect:param} allows us to visualise the bispectrum
dependence on the configurations. The bispectrum is computed for 32
perimeters in the $\ell$ space. Only nine perimeters are shown for
illustration in Figs. \ref{fig:paramrad} and \ref{fig:paramir}, for
the radio and IR-source populations respectively. The bispectrum
values are colour coded from blue (lowest value) to red (highest
value). The succession of plots, arranged by increasing perimeters,
exhibits the allowed configurations at given perimeter with the
equilateral configuration being the starting (upper left panel) and
ending point (lower right panel). Unsurprisingly the bispectrum
amplitude of the radio sources does not vary with the configuration
(same color code for all points in Fig. \ref{fig:paramrad}). As for
the IR sources, Fig. \ref{fig:paramir}, the bispectrum amplitude
decreases with perimeter, thus from upper left to lower right
panel. Moreover, it is worth noting that the bispectrum values do not
vary vertically. This means that within the proposed $(P, F,
S)$-parameterisation the bispectrum of the IR sources is
quasi-independent of $F$, reducing the full bispectrum to a function
of the two parameters $P$ and $S$.  Finally, at a given perimeter,
i.e. scale, the bispectrum of the IR sources is more dependent on the
configuration and peaks in the squeezed triangles, upper left points
in panels 2 to 5.

\subsection{point sources contamination of $f_\mathrm{NL}$}

We now explore the point-source non-Gaussianity in terms of
contamination of the $f_\mathrm{NL}$ estimation.  For pedagogical
purposes, we consider the SW regime, i.e. a constant transfer function
without acoustic oscillations, damping etc.

For IR sources alone, which dominate at high frequencies,
$b(\ell_1,\ell_2,\ell_3)\propto\sqrt{C_{\ell_1}C_{\ell_2}C_{\ell_3}}$. Combined
with Planck's latest constraints on the CIB $\ell \!\times\! C_\ell
\!\!\simeq \!\!\mathrm{const}$ \citep{Planck-Collaboration-CIB}, this
yields:
\begin{equation*}
b^\mathrm{IR}(\ell_1,\ell_2,\ell_3) \propto
\frac{1}{\sqrt{\ell_1\ell_2\ell_3}}
\end{equation*}
which has a similar shape to the local template in the SW limit,
Eq. (\ref{eq:cmbispsw}).
We define the scalar product between the bispectra $f,g$:
\begin{equation*}
<f,g>=\sum_{\ell_1\leq\ell_2\leq\ell_3} N_{\ell_1 \ell_2 \ell_3} \frac{f(\ell_1,\ell_2,\ell_3)
  \times g(\ell_1,\ell_2,\ell_3)}{C_{\ell_1}^{\mathrm{CMB}} \,
  C_{\ell_2}^{\mathrm{CMB}} \, C_{\ell_3}^{\mathrm{CMB}}} 
\end{equation*}
where the denominator is the variance of the local bispectrum, for
triangles with $\ell_1 \neq \ell_2 \neq \ell_3$.  The correlation
coefficient between a bispectrum $b^\alpha$ and the local bispectrum
$b^\mathrm{loc}$ is:
\begin{equation*}
\cos \theta_{\alpha} = \frac{<b^{\alpha},b^\mathrm{loc}>}{\|
  b^{\alpha}\| \; \| b^\mathrm{loc}\|} 
\end{equation*}
with $\alpha$ being IR or radio.

\begin{figure}
\centering
\includegraphics[width=8.5cm]{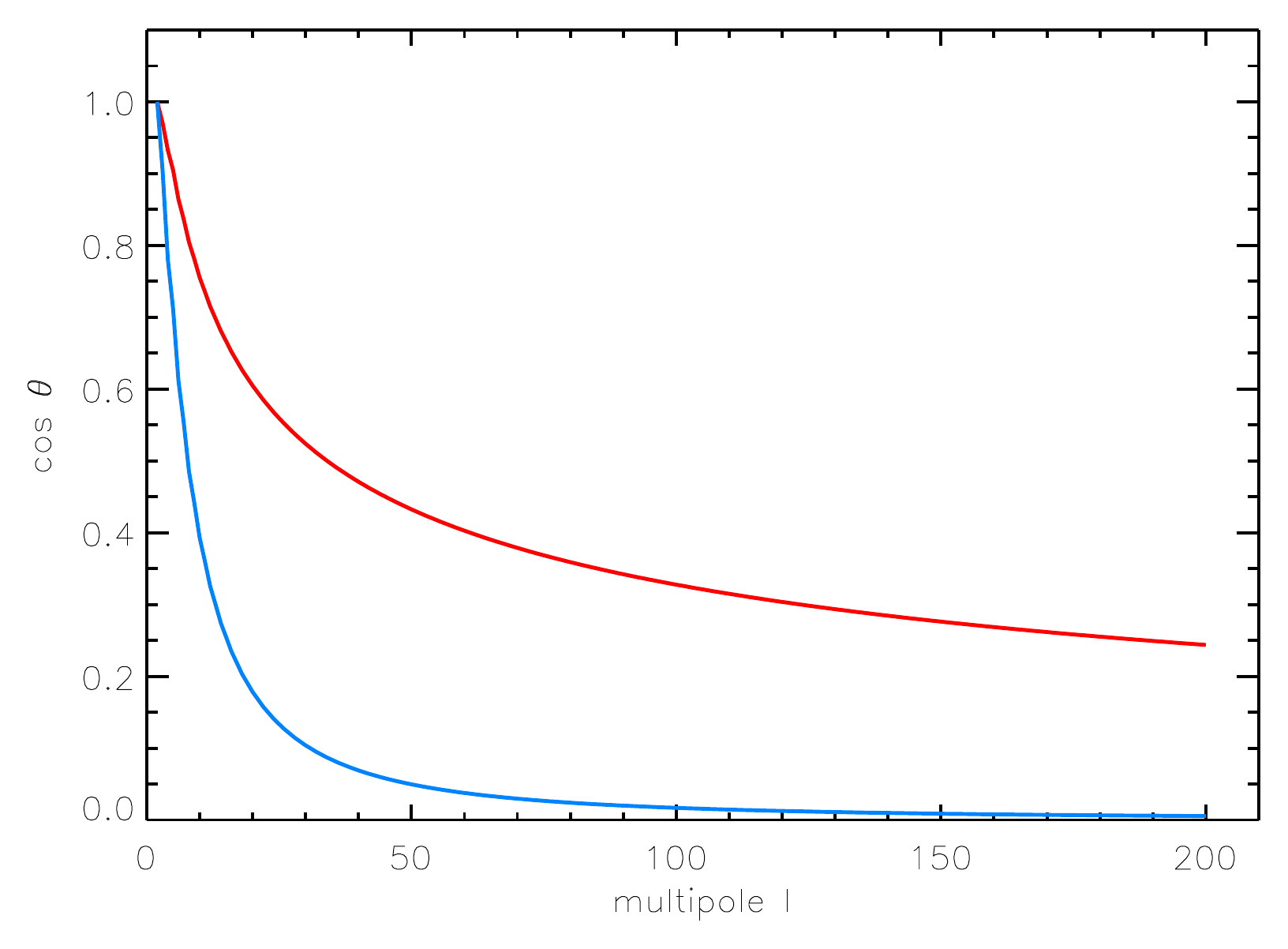}
\caption{Correlation between SW local bispectrum and respectively IR
  (red) and radio (blue) bispectrum, as a function of the maximum
  multipole used ($\ell_\mathrm{min}=2$)} 
\label{costhetafg}
\end{figure}

Figure \ref{costhetafg} shows that, for $\ell<200$, the correlation
between radio and local bispectra decreases quickly, so that the two
bispectra may be distinguished efficiently. Conversely, even when
using a large multipole range, the IR bispectrum is significantly
correlated with the local one.

The contribution of the bispectrum of a point source population
$\alpha$ to $f_\mathrm{NL}$ is: 
\begin{equation}\label{eq:dfnlalpha}
\Delta f_\mathrm{NL}^{\alpha} =
\frac{<b^{\alpha},b^{f_\mathrm{NL}=1}>}{<b^{f_\mathrm{NL}=1},b^{f_\mathrm{NL}=1}>}
= \frac{\| b^{\alpha}\|}{ \| b^{f_\mathrm{NL}=1}\|} \, \cos
\theta_{\alpha}
\end{equation}
This equation is the usual bias \citep{Serra2008} of the
local-optimised NG estimator, when the local bispectrum has the form
of Eq. (\ref{eq:cmbispsw}).

A more comprehensive computation of $\Delta f_\mathrm{NL}^{\alpha}$ is
achievable by applying the full local estimator described in Sect.
\ref{sect:fnlestim}. We built up this estimator using the full
transfer function from the latest version of CAMB \citep{Lewis2000}
with WMAP7+BAO+H0 cosmological parameters \citep{Larson2011}, and we
tested the estimator on simulations from \cite{Elsner2009}. We found
the previously noted result that the variance of the estimator increases
with $f_\mathrm{NL}$. It is unbiased in the range we have tested ($0
\leq f_\mathrm{NL} \leq 200$).

We used this $f_\mathrm{NL}$ estimator on two sets of simulated maps:
maps containing all the point sources, and maps with only sources
below the flux limit of Planck's Early Release Compact Source
Catalogue (ERCSC) \citep{Planck-Collaboration-ERCSC}, namely $S_c
=$0.5, 0.5, 0.3, 0.3, 0.3, 0.25 Jy as a function of
frequency. Moreover, we have computed the estimator at three
resolutions, $\ell_\mathrm{max}$, recalibrating the
$\mathrm{S}_\mathrm{prim}$ normalisation in each case.  Tables
\ref{fnlradtable} and \ref{fnlirtable} summarise these results.

\begin{table}
\centering
\begin{tabular}{@{}l@{}|@{}c@{}c@{}c@{}c@{}c@{}c@{}}
\hline
\multicolumn{7}{c}{without flux cut}\\
\hline
\hline
 $\nu$ (GHz) & $\ $30$\ $ & 90$\ $ & 148$\ $ & 219$\ $  & 277$\ $ & 350\\
\hline
$\ell_\mathrm{max}=50$ & -4.2& -0.0025 & -0.00037 & -0.00021 & -0.00027 & -0.00068\\
$\ell_\mathrm{max}=700$ & 3850& 2.5 & 0.38 & 0.21 & 0.27 & 0.65\\
$\ell_\mathrm{max}=2048\ $ & 177000& 117& 18& 9.7 & 12& 30 \\
\hline
\hline
\multicolumn{7}{c}{with flux cut}\\
\hline
\hline
$\nu$ (GHz) & $\ $30$\ $ & 90$\ $ & 148$\ $ & 219$\ $  & 277$\ $ & 350\\
\hline
$\ell_\mathrm{max}=700$ & 108 &Ê0.17Ê& 0.0071Ê& 0.0031Ê& 0.0035Ê& 0.0064\\
$\ell_\mathrm{max}=2048\ $ & 4930Ê& 7.5Ê& 0.31Ê& 0.14 & 0.16Ê& 0.29 \\
\hline
\hline
\end{tabular}
\caption{Bias on the $f_\mathrm{NL}$ estimator, $\Delta
  f_\mathrm{NL}^\mathrm{RAD}$, due to radio sources}
\label{fnlradtable}
\end{table}

\begin{table}
\centering
\begin{tabular}{@{}l@{}|@{}c@{}c@{}c@{}c@{}c@{}c@{}}
\hline
\multicolumn{7}{c}{without flux cut}\\
\hline
\hline
 $\nu$ (GHz) & 30 & 90 & 148 & 219$\ $  & 277$\ $ & 350\\
\hline
$\ell_\mathrm{max}=50$ & $\,$-$3.5 \!\cdot\! 10^{-8}$ & $\,$-$5.9 \!\cdot\! 10^{-6}$ & $\,$-$9.2 \!\cdot\! 10^{-5}$ & -0.0027 & -0.023 & -1.0\\
$\ell_\mathrm{max}=700$ & $\,$-$1.3 \!\cdot\! 10^{-6}$ & -0.00019 & -0.0033 & -0.063 & -0.55 & -9.0 \\
$\ell_\mathrm{max}=2048\,$ & $\,$-$1.8 \!\cdot\! 10^{-5}$& -0.0026 & -0.039 & -0.68 & -4.8 & -67\\
\hline
\hline
\multicolumn{7}{c}{with flux cut}\\
\hline
\hline
$\nu$ (GHz) & $\ $30 & 90 & 148 & 219$\ $  & 277$\ $ & 350\\
\hline
$\ell_\mathrm{max}=700$ & $\,$-$1.3 \!\cdot\! 10^{-6}$ & -0.00019 & -0.0033 & -0.078 & -0.74 & -11\\
$\ell_\mathrm{max}=2048\,$ &  $\,$-$1.8 \!\cdot\! 10^{-5}$& -0.0026 & -0.039 & -0.67 & -6.3 & -66 \\
\hline
\hline
\end{tabular}
\caption{Bias of the $f_\mathrm{NL}$ estimator, $\Delta
  f_\mathrm{NL}^\mathrm{IR}$, due to IR sources}
\label{fnlirtable}
\end{table}

The bias $\Delta f_\mathrm{NL}^\mathrm{RAD}$ (see Table
\ref{fnlradtable}) is negative on large angular scales, for
$\ell_\mathrm{max}=50$. The bias due to radio sources becomes positive
at higher multipoles in agreement with \cite{Serra2008}. This is due
to the CMB bispectrum being negative in the SW-dominated regime and
the radio bispectrum being positive. The bias increases by 5 orders of
magnitude at the highest resolution, $\ell_\mathrm{max}=2048$. The
reason for the rapid increase of the bias with $\ell_{\mathrm{max}}$
relates to the weight of the observed bispectrum in
Eq. (\ref{eq:Dfnlanalyt}), $B^\mathrm{loc}_{\ell_1\ell_2\ell_3} /
C_{\ell_1}C_{\ell_2}C_{\ell_3}$, which rapidly increases with
multipole as the product of spectra decreases more quickly than the
bispectrum. This leads to a $1/C_\ell$ dependence in squeezed
configurations and to a $1/C_\ell^2$ dependence in equilateral
configurations.  When the observed bispectrum is associated to CMB
signal alone, its decrease cancels the increase of the weights so that
the sum in Eq. (\ref{eq:Dfnlanalyt}) converges. Conversely, the sum
diverges when the observed bispectrum is associated with a non-CMB
signal and does not decrease with $\ell$ as fast as the CMB.

The bias $\Delta f_\mathrm{NL}^\mathrm{RAD}$ is maximal at 30 GHz and
rapidly decreases with frequency. It slightly increases again at the
two highest frequencies following the amplitude of the bispectrum in
temperature units which is plotted in the upper panel of
Fig. \ref{fig:amplRAD}. The relative error of $\Delta
f_\mathrm{NL}^\mathrm{RAD}$ for $\ell_{\mathrm{max}}=700$ is of the
order of 1.5\% independently of the frequency. It amounts to 2.3\% for
$\ell_{\mathrm{max}}=2048$. These errors bars were computed with
simulations using the catalog of sources present in Sehgal et al.'s
maps

As shown in Table \ref{fnlradtable}, masking sources above the ERCSC
flux limit proves very efficient to significantly decrease the radio
contamination to $f_\mathrm{NL}$ at all the frequencies. At a
Planck-like resolution, $\ell_\mathrm{max}= 2048$, the bias $\Delta
f_\mathrm{NL}^\mathrm{RAD}$ is reduced below unity above 150 GHz. It
is of the order of Planck's expected error bars at 90 GHz. At 30 GHz
the bias is still important.

The bias due to IR sources $\Delta f_\mathrm{NL}^\mathrm{IR}$ is
always negative, see Table \ref{fnlirtable}. As a matter of fact, we
have shown that the IR bispectrum peaks in squeezed configurations
just like the CMB bispectrum and these configurations thus dominate
the sum in Eq. (\ref{eq:Dfnlanalyt}). Moreover, in the squeezed limit
the CMB bispectrum is negative while the IR bispectrum is
positive. For the same reason as for radio sources, the bias $\Delta
f_\mathrm{NL}^\mathrm{IR}$ blows up at high multipoles. This is
particularly important at a Planck-like resolution,
$\ell_\mathrm{max}=2048$, where primordial NG tests will need to
carefully handle the contamination by IR sources. The IR sources
emission plummets at radio frequencies so that $\Delta
f_\mathrm{NL}^\mathrm{IR}$ is completely negligible below 220 GHz.  It
becomes of the order of Planck's error bars at 277 GHz and it reaches
WMAP's central values for $f_\mathrm{NL}$ at 350 GHz. The relative
error of $\Delta f_\mathrm{NL}^\mathrm{IR}$ ranges between 6 and 7\%
from 148 to 350 GHz for $\ell_{\mathrm{max}}=700$. It ranges between 3
and 7\% for $\ell_{\mathrm{max}}=2048$. (These error bars were
computed analytically with the weak NG approximation -- see Appendix
\ref{appendix:wngvar}) At higher frequencies the IR contamination to
the bispectrum is likely larger but the contamination from our Galaxy
needs to be taken into account as well.

Interestingly, masking sources above the ERCSC flux limit does not
diminish $\Delta f_\mathrm{NL}^\mathrm{IR}$, as most of the IR sources
are unresolved and the IR clustering is mostly due to faint
sources. Masking may even artificially boost $\Delta
f_\mathrm{NL}^\mathrm{IR}$, for example at 277 GHz, since it mostly
affects the flat shot-noise which produces a positive bias $\Delta
f_\mathrm{NL}$.


\section{Conclusions and discussion} \label{sect:concl}

We have studied the non-Gaussianity produced by point sources in the
frequency range of the CMB from 30 to 350 GHz. We have developed a
simple and accurate prescription to infer the angular bispectrum from the
power spectrum of point sources, considering different independent populations
of sources, with or without clustering.

Using publicly available all-sky simulations of radio and IR sources, we have
computed the full-sky binned bispectra for these two populations of
sources. We have compared the measured bispectra to those predicted
from our prescription and found a very good agreement between the two.
We have displayed the angular bispectrum using a new parametrisation
which highlights efficiently the configuration dependence.  

We have characterised the angular bispectrum of the IR and
  radio sources showing the configuration dependence and the
frequency behaviour. In particular and for the first time, we showed
that the IR bispectrum peaks in the squeezed triangles and that the
clustering of IR sources enhances the bispectrum values by several
orders of magnitude on large angular scales $\ell \sim 100$.  The
bispectrum of IR sources starts to dominate that of radio sources
on large angular scales at 150 GHz, and it dominates the whole
multipole range at 350 GHz.

Finally to illustrate the contamination of local CMB non-Gaussianity
by point sources, we derive the bias on $f_\mathrm{NL}$ induced by
radio and IR sources, for WMAP or Planck-like angular
resolutions. Radio sources produce a positive bias which is
significantly reduced ($\Delta f_\mathrm{NL} < 1$ for $\nu \geq$ 150
GHz) by masking the sources above a given flux limit taken as the
ERCSC cut. The form of the IR bispectrum mimics a primordial `local'
bispectrum $f_\mathrm{NL}$ on large angular scales. The IR sources
produce a negative bias which becomes important for Planck-like
resolution and at high frequencies ($\Delta f_\mathrm{NL} \sim -6$ at
277 GHz and $\Delta f_\mathrm{NL} \sim -$60-70 at 350 GHz). Most of
the signal is associated with the clustering of faint IR
sources. Therefore, the bias $\Delta f_\mathrm{NL}^\mathrm{IR}$ is not
reduced by masking sources above a flux limit but, in some cases, even
increased due to the reduction of the shot-noise term.

Our analysis highlights the sensitivity of the bias on $f_\mathrm{NL}$
to the experimental properties (maximum resolution and frequency range),
the point-source models (clustering or no, resolved or not) and
their scale dependence with respect to the CMB. For high resolution
high frequency CMB experiments, primordial NG estimations need to take
special care of astrophysical contaminations. One solution would be to 
estimate the primordial and astrophysical non-Gaussianity simultaneously.


\section*{Acknowledgments}
The authors thank an anonymous referee for comments and
  suggestions. They wish to thank S. Ilic, G. Lagache and A. Penin
for useful discussions. They acknowledge the use of Lambda
archive\footnote{http://lambda.gsfc.nasa.gov/toolbox/tb\_cmbsim\_ov.cfm},
CAMB \citep{Lewis2000} and the HEALPix \citep{Gorski:2004by}
package. They made use of all-sky simulations of the microwave sky by
\cite{Sehgal2010} and non-Gaussian CMB simulations by
\cite{Elsner2009}. NA and FL thank Universit\'{e} de Gen\`{e}ve
and the Swiss NSF for partial support and hosting, MK and MF thank the
IAS for hospitality on numerous occasions. The authors acknowledge
partial support from PHC Germaine de Sta\"{e}l. FL further
acknowledges financial support from a PhD fellowship of the Ecole
Normale Sup\'{e}rieure Paris. MK and MF acknowledge funding by the
Swiss NSF.  Part of the calculations were performed on
the {\em Andromeda} cluster of the Universit\'{e} de Gen\`{e}ve.


\appendix
\section{Clustered sources shot-noise}\label{appendix:shot}
A source with flux $S$ enclosed in a pixel $\Omega_{\mathrm{pix}}$
yields a rise of temperature compared to the CMB: 
\begin{equation}
\Delta T = \frac{S}{\Omega_{\mathrm{pix}}} \times
\underbrace{\frac{(e^x-1)^2}{x^2 e^x} \times
  \frac{c^2}{2\nu^2 k_B}}_{\equiv k_\nu} 
\end{equation}
where $x=h\nu/k_\mathrm{B} T_\mathrm{CMB}$ and $k_\nu = \left.\frac{\partial
  B(\nu,T)}{\partial T}\right|_{T_\mathrm{CMB}}$. 
  
The two-point correlation function of point sources takes the form:
\begin{equation*}
\langle \Delta T(\mathbf{n}) \Delta T(\mathbf{n}') \rangle  =
F(\mathbf{n},\mathbf{n}') + \Gamma \, \delta_{\mathbf{n},\mathbf{n}'} 
\end{equation*}
where $F(\mathbf{n},\mathbf{n}')$ is the correlation function coming
from the spatial 
distribution of the sources, and the Kronecker term comes from the
discreteness of the sources:
\begin{equation*}
\Gamma  =  \langle\Delta T^2\rangle - F(\mathbf{n},\mathbf{n})
\end{equation*}
Assuming statistical isotropy, we get:
\begin{equation*}
C_\ell = C_\ell^\mathrm{clust} + C_\ell^{\mathrm{shot}} \quad
\mathrm{with} \quad C_\ell^{\mathrm{shot}} = \Gamma
\,\Omega_{\mathrm{pix}} 
\end{equation*}
Indeed:
\begin{multline}
\langle a_{\ell m} \, a^*_{\ell' m'}\rangle \, = \!\int \! \mathrm{d}^2\mathbf{n} \, \mathrm{d}^2\mathbf{n}' \, Y_{\ell m}(\mathbf{n}) \, Y^*_{\ell' m'}(\mathbf{n}') \langle \Delta T(\mathbf{n}) \Delta T(\mathbf{n}') \rangle \\ 
= \underbrace{\int \! \mathrm{d}^2\mathbf{n} \, \mathrm{d}^2\mathbf{n}' \, Y_{\ell m}(\mathbf{n}) \, Y^*_{\ell' m'}(\mathbf{n}') \, F(\mathbf{n},\mathbf{n}')}_{ = \, C_\ell^\mathrm{clust} \, \delta_{\ell \ell'} \, \delta_{m m'}} \\
+ \sum_{\mathbf{n}_i,\mathbf{n}'_j} \, Y_{\ell m}(\mathbf{n}_i) \, Y^*_{\ell' m'}(\mathbf{n}'_j) \!\times\! \Gamma \, \delta_{\mathbf{n}_i,\mathbf{n}'_j} \, \Omega^2_{\mathrm{pix}}
\\= C_\ell^\mathrm{clust} \, \delta_{\ell \ell'} \, \delta_{m m'} + \Gamma \, \Omega_{\mathrm{pix}} \sum_{\mathbf{n}_i} Y_{\ell m}(\mathbf{n}_i) \, Y^*_{\ell' m'}(\mathbf{n}_i)\, \Omega_{\mathrm{pix}}
\\=C_\ell^\mathrm{clust} \, \delta_{\ell \ell'} \, \delta_{m m'} +
\Gamma \, \Omega_{\mathrm{pix}} \, \int \mathrm{d}^2\mathbf{n}
\,Y_{\ell m}(\mathbf{n}) \, Y^*_{\ell' m'}(\mathbf{n}) 
\\=C_\ell^\mathrm{clust} \, \delta_{\ell \ell'} \, \delta_{m m'}
+\Gamma \, \Omega_{\mathrm{pix}} \, \delta_{\ell \ell'} \, \delta_{m
  m'} 
\end{multline}
$C_\ell$ has units $\mu K^2\cdot \mathrm{sr}\,$. Let us number by
i=1..N all sources of the sky, then the temperature of a pixel is
given by: 
\begin{equation}
\Delta T(\mathbf{n})= \frac{k_\nu}{\Omega_{\mathrm{pix}}} \sum_{i=1}^N
S_i \times \mathds{1}_{[i \in \mathbf{n}]} 
\end{equation}
where $\mathds{1}_{[i \in \mathbf{n}]}$ is 1 if the source i is in the
pixel and 0 otherwise. We have: 
$$\langle \mathds{1}_{[i \in \mathbf{n}]}\rangle =1/n_\mathrm{pix}$$
Hence for $\mathbf{n} \neq \mathbf{n}'$:
\begin{eqnarray*}
F(\mathbf{n},\mathbf{n}') & = & \langle \Delta T(\mathbf{n}) \Delta
T(\mathbf{n}') \rangle\\ 
& = & \frac{k^2_\nu}{\Omega^2_{\mathrm{pix}}} \langle \sum_{i\neq j} S_i \,
S_j \times \mathds{1}_{[i \in \mathbf{n}]} \mathds{1}_{[j \in
    \mathbf{n}']}\rangle 
\end{eqnarray*}
Then we find:
\begin{eqnarray*}
\langle \Delta T(\mathbf{n})^2\rangle & = & \frac{k^2_\nu}{\Omega^2_{\mathrm{pix}}} \langle \sum_{i,j=1}^N S_i \, S_j \times \mathds{1}_{[i \in \mathbf{n}]} \mathds{1}_{[j \in \mathbf{n}]}\rangle\\
& = & \frac{k^2_\nu}{\Omega^2_{\mathrm{pix}}} \langle\sum_{i\neq j} S_i \, S_j \times \mathds{1}_{[i \in \mathbf{n}]} \mathds{1}_{[j \in \mathbf{n}]}\rangle \\
& & + \frac{k^2_\nu}{\Omega^2_{\mathrm{pix}}} \sum_{i=1}^N S^2_i \times \langle\mathds{1}_{[i \in \mathbf{n}]}\rangle\\
& = & \lim_{\mathbf{n}\rightarrow \mathbf{n}'} \langle\Delta T(\mathbf{n}) \Delta T(\mathbf{n}')\rangle + \frac{k^2_\nu}{\Omega^2_{\mathrm{pix}}} \frac{1}{n_{\mathrm{pix}}} \sum_{\mathrm{sources}} S^2
\end{eqnarray*}
Recalling $\Omega_{\mathrm{pix}}=\frac{4\pi}{n_{\mathrm{pix}}}$ and introducing $\frac{dn}{dS}$ the number counts of sources
\begin{eqnarray*}
\langle\Delta T(\mathbf{n})^2\rangle & = & F(\mathbf{n},\mathbf{n}) + \underbrace{\frac{k^2_\nu}{4\pi \,\Omega_{\mathrm{pix}}} \int S^2 \, \frac{\mathrm{d}n}{\mathrm{d}S} \, \mathrm{d}S}_{=\,\Gamma}
\end{eqnarray*}
And finally:
\begin{equation}
C_\ell^\mathrm{shot} = \Gamma \,\Omega_{\mathrm{pix}} = \frac{k^2_\nu}{4\pi} \int S^2 \, \frac{\mathrm{d}n}{\mathrm{d}S} \, \mathrm{d}S
\end{equation}
which is the shot-noise formula Eq. (\ref{shotspec}).
\\The integral runs from S=0 to $S_\mathrm{cut}$ the flux detection limit of the survey, ie
sources with $S>S_\mathrm{cut}$ have been removed. Note that this result is independent of the two-point correlation function, which we did not specify.
\\At order 3 for the angular bispectrum, the computation is a bit more involved but follows the same line, and we find:
\begin{equation}
\langle a_{\ell_1 m_1} a_{\ell_2 m_2} a_{\ell_3 m_3} \rangle =
G_{\ell_1 \ell_2 \ell_3}^{m_1 m_2 m_3} \times b_{\ell_1 \ell_2 \ell_3}
\end{equation}
with
\begin{equation}
b^{\mathrm{shot}}_{\ell_1\ell_2\ell_3}  =  \frac{k_\nu^3}{4\pi} \times \int S^3 \,\frac{\mathrm{d}n}{\mathrm{d}S} \, \mathrm{d}S
\end{equation}
which is the shot-noise formula Eq. (\ref{shotbisp}).

\section{Bispectrum variance in the weak NG
  approximation}\label{appendix:wngvar} 

The bispectrum estimator Eq. (\ref{eq:estimbisp}) can be put in the form:
\begin{equation}
\hat{b}_{123} = \frac{1}{N_{123}} \times \sum_{m_{123}} G_{123} \, a_1
\, a_2 \, a_3
\end{equation}
where a shortened notation is used: $N_{123}$ is the number of
triangles defined in Eq. \ref{eq:ntri}, and $G_{123}$ is the Gaunt
coefficient:
\begin{equation}
G_{123} = \int \mathrm{d}^2 \mathbf{n} \, Y_{\ell_1 m_1}(\mathbf{n})
Y_{\ell_2 m_2}(\mathbf{n}) Y_{\ell_3 m_3}(\mathbf{n}) \,.
\end{equation}
Then the bispectrum covariance takes the form
\begin{eqnarray}
\nonumber\mathrm{Cov}(\hat{b}_{123}\, , \hat{b}_{1'2'3'}) & = & \frac{1}{N_{123} \, N_{1'2'3'}} \sum_{m_{123},m'_{123}} G_{123} \,G_{1'2'3'}\\
&& \!\!\!\!\!\!\!\!\!\!\!\!\!\!\!\!\!\!\!\!\!\!\!\!\left(\langle a_1 a_2 a_3 a_{1'} a_{2'} a_{3'}\rangle - \langle a_1 a_2 a_3\rangle\langle a_{1'} a_{2'} a_{3'}\rangle\right)\qquad
\end{eqnarray}
When the field is close to Gaussian, the 6-point correlation function
can be computed with Wick's theorem (\cite{Wick1950}, \cite{Komatsu2002} and references therein): 
\begin{equation}
\langle a_1 a_2 a_3 a_{1'} a_{2'} a_{3'}\rangle = (C_\ell)_{123} \,\delta_{1^*,1'} \,\delta_{2^*,2'} \,\delta_{3^*,3'} +14\,\mathrm{perm.} 
\end{equation}
where $\delta_{i^*,j} = (-1)^{m_i} \delta_{\ell_i,\ell_j} \delta_{-m_i,m_j}$.\\
The 15 permutations of (1,2,3,1',2',3') are listed below along with their contribution $\delta\mathrm{Cov}$ to the covariance.
\begin{eqnarray}
(1^*2)(3^*1')(2'^*3') \rightarrow \delta\mathrm{Cov}=&0 \mathrm{\ except \ if \ \ell_3=\ell'_1=0}\\
(1^*2)(3^*2')(1'^*3') \rightarrow \delta\mathrm{Cov}=&0 \mathrm{\ except \ if \ \ell_3=\ell'_2=0}\\
(1^*2)(3^*3')(1'^*2') \rightarrow \delta\mathrm{Cov}=&0 \mathrm{\ except \ if \ \ell_3=\ell'_3=0} \\
(1^*3)(2^*1')(2'^*3') \rightarrow \delta\mathrm{Cov}=&0 \mathrm{\ except \ if \ \ell_2=\ell'_1=0} \\
(1^*3)(2^*2')(1'^*3') \rightarrow \delta\mathrm{Cov}=&0 \mathrm{\ except \ if \ \ell_2=\ell'_2=0} \\
(1^*3)(2^*3')(1'^*2') \rightarrow \delta\mathrm{Cov}=&0 \mathrm{\ except \ if \ \ell_2=\ell'_3=0} \\
(1^*1')(2^*3)(2'^*3') \rightarrow \delta\mathrm{Cov}=&0 \mathrm{\ except \ if \ \ell_1=\ell'_1=0} \\
(1^*1')(2^*2')(3^*3') \rightarrow \delta\mathrm{Cov}=&\!\!\!\!\frac{C_{\ell_1}C_{\ell_2}C_{\ell_3}}{N_{123}} \delta_{\ell_1\ell'_1}\;\!\delta_{\ell_2\ell'_2}\;\!\delta_{\ell_3\ell'_3}\\
(1^*1')(2^*3')(3^*2') \rightarrow \delta\mathrm{Cov}=&\!\!\!\!\frac{C_{\ell_1}C_{\ell_2}C_{\ell_3}}{N_{123}} \delta_{\ell_1\ell'_1}\;\!\delta_{\ell_2\ell'_3}\;\!\delta_{\ell_3\ell'_2}\\
(1^*2')(2^*3)(1'^*3') \rightarrow \delta\mathrm{Cov}=&0 \mathrm{\ except \ if \ \ell_1=\ell'_2=0} \\
(1^*2')(2^*1')(3^*3') \rightarrow \delta\mathrm{Cov}=&\!\!\!\!\frac{C_{\ell_1}C_{\ell_2}C_{\ell_3}}{N_{123}} \delta_{\ell_1\ell'_2}\;\!\delta_{\ell_2\ell'_1}\;\!\delta_{\ell_3\ell'_3}\\
(1^*2')(2^*3')(3^*1') \rightarrow \delta\mathrm{Cov}=&\!\!\!\!\frac{C_{\ell_1}C_{\ell_2}C_{\ell_3}}{N_{123}} \delta_{\ell_1\ell'_2}\;\!\delta_{\ell_2\ell'_3}\;\!\delta_{\ell_3\ell'_1} \\
(1^*3')(2^*3)(1'^*2') \rightarrow \delta\mathrm{Cov}=&0 \mathrm{\ except \ if \ \ell_1=\ell'_3=0} \\
(1^*3')(2^*1')(3^*2') \rightarrow \delta\mathrm{Cov}=&\!\!\!\!\frac{C_{\ell_1}C_{\ell_2}C_{\ell_3}}{N_{123}} \delta_{\ell_1\ell'_3}\;\!\delta_{\ell_2\ell'_1}\;\!\delta_{\ell_3\ell'_2} \\
(1^*3')(2^*2')(3^*1') \rightarrow \delta\mathrm{Cov}=&\!\!\!\!\frac{C_{\ell_1}C_{\ell_2}C_{\ell_3}}{N_{123}} \delta_{\ell_1\ell'_3}\;\!\delta_{\ell_2\ell'_2}\;\!\delta_{\ell_3\ell'_1} 
\end{eqnarray}
Here we do not consider bispectrum coefficients with one multipole
equal to zero (which amounts to considering the power spectrum times
the monopole). So the bispectrum covariance is diagonal and we find :
\begin{equation}
\mathrm{Var}(\hat{b}_{123}) =
\frac{C_{\ell_1}C_{\ell_2}C_{\ell_3}}{N_{123}} \times
\left\{ \begin{array}{ll} 6 & \mathrm{equilateral \ triangle}\\ 2 &
  \mathrm{isosceles \ triangle}\\ 1 & \mathrm{general
    \ triangle} \end{array}\right.
\end{equation}


\bibliographystyle{mn2e}
\bibliography{article}

\begin{thebibliography}{}

\bibitem[\protect\citeauthoryear{Acquaviva, Bartolo, Matarrese \&
  Riotto}{Acquaviva et~al.}{2003}]{Acquaviva2003}
Acquaviva V.,  Bartolo N.,  Matarrese S.,    Riotto A.,  2003, Nuclear Physics
  B, 667, 119

\bibitem[\protect\citeauthoryear{Aghanim, Kunz, Castro \& Forni}{Aghanim
  et~al.}{2003}]{Aghanim:2003fs}
Aghanim N.,  Kunz M.,  Castro P.,    Forni O.,  2003, Astron.Astrophys., 406,
  797

\bibitem[\protect\citeauthoryear{Aghanim, Majumdar \& Silk}{Aghanim
  et~al.}{2008}]{Aghanim2008}
Aghanim N.,  Majumdar S.,    Silk J.,  2008, Reports on Progress in Physics,
  71, 066902

\bibitem[\protect\citeauthoryear{Amblard, Cooray, Serra \& et al.}{Amblard
  et~al.}{2011}]{Amblard2011}
Amblard A.,  Cooray A.,  Serra P.,    et al. 2011, Nature, 470, 510

\bibitem[\protect\citeauthoryear{Argueso, GonzalezâNuevo \& Toffolatti}{Argueso
  et~al.}{2003}]{Argueso2003}
Argueso F.,  GonzalezâNuevo J.,    Toffolatti L.,  2003, The Astrophysical
  Journal, 598, 86

\bibitem[\protect\citeauthoryear{{Astier}, {Guy}, {Regnault}, {Pain},
  {Aubourg}, {Balam}, {Basa}, {Carlberg}, {Fabbro}, {Fouchez}, {Hook}
  et~al.,}{{Astier} et~al.}{2006}]{Astier2006}
{Astier} P.,  {Guy} J.,  {Regnault} N.,  {Pain} R.,  {Aubourg} E.,  {Balam} D.,
   {Basa} S.,  {Carlberg} R.~G.,  {Fabbro} S.,  {Fouchez} D.,  {Hook} I.~M.,
  et~al., 2006, Astronomy and Astrophysics, 447, 31

\bibitem[\protect\citeauthoryear{Babich \& Pierpaoli}{Babich \&
  Pierpaoli}{2008}]{Babich2008}
Babich D.,  Pierpaoli E.,  2008, Physical Review D, 77, 123011

\bibitem[\protect\citeauthoryear{{Bartolo}, {Komatsu}, {Matarrese} \&
  {Riotto}}{{Bartolo} et~al.}{2004}]{Bartolo2004}
{Bartolo} N.,  {Komatsu} E.,  {Matarrese} S.,    {Riotto} A.,  2004, Physics
  Reports, 402, 103

\bibitem[\protect\citeauthoryear{{Bassett}, {Tsujikawa} \& {Wands}}{{Bassett}
  et~al.}{2006}]{Bassett2006}
{Bassett} B.~A.,  {Tsujikawa} S.,    {Wands} D.,  2006, Reviews of Modern
  Physics, 78, 537

\bibitem[\protect\citeauthoryear{{Blake}, {Davis}, {Poole}, {Parkinson},
  {Brough} et~al.,}{{Blake} et~al.}{2011}]{Blake2011}
{Blake} C.,  {Davis} T.,  {Poole} G.~B.,  {Parkinson} D.,  {Brough} S.,
  et~al., 2011, Mon.Not.Roy.Astron.Soc., 415, 2892

\bibitem[\protect\citeauthoryear{{Boughn} \& {Partridge}}{{Boughn} \&
  {Partridge}}{2008}]{Boughn2008}
{Boughn} S.~P.,  {Partridge} R.~B.,  2008, Publications of the Astronomical
  Society of the Pacific, 120, 281

\bibitem[\protect\citeauthoryear{Bucher, Tent \& Carvalho}{Bucher
  et~al.}{2010}]{Bucher2010}
Bucher M.,  Tent B.~V.,    Carvalho C.~S.,  2010, Monthly Notices of the Royal
  Astronomical Society, 407, 2193

\bibitem[\protect\citeauthoryear{Byrnes \& Choi}{Byrnes \&
  Choi}{2010}]{Byrnes2010}
Byrnes C.~T.,  Choi K.-Y.,  2010, Advances in Astronomy, 2010, 1

\bibitem[\protect\citeauthoryear{Cooray \& Kesden}{Cooray \&
  Kesden}{2002}]{Cooray2002}
Cooray A.,  Kesden M.,  2002, New Astronomy, 8, 21

\bibitem[\protect\citeauthoryear{Creminelli, Nicolis, Senatore, Tegmark \&
  Zaldarriaga}{Creminelli et~al.}{2006}]{Creminelli2006}
Creminelli P.,  Nicolis A.,  Senatore L.,  Tegmark M.,    Zaldarriaga M.,
  2006, Journal of Cosmology and Astroparticle Physics, 2006, 004

\bibitem[\protect\citeauthoryear{Creminelli \& Zaldarriaga}{Creminelli \&
  Zaldarriaga}{2004}]{Creminelli2004}
Creminelli P.,  Zaldarriaga M.,  2004, Journal of Cosmology and Astroparticle
  Physics, 2004, 006

\bibitem[\protect\citeauthoryear{{Das}, {Marriage}, {Ade}, {Aguirre}, {Amiri},
  {Appel}, {Barrientos} \& et al.}{{Das} et~al.}{2011}]{Das2011}
{Das} S.,  {Marriage} T.~A.,  {Ade} P.~A.~R.,  {Aguirre} P.,  {Amiri} M.,
  {Appel} J.~W.,  {Barrientos} L.~F.,    et al. 2011, The Astrophysical
  Journal, 729, 62

\bibitem[\protect\citeauthoryear{De~Troia, Ade, Bock, Bond, Boscaleri
  et~al.,}{De~Troia et~al.}{2003}]{DeTroia:2003tq}
De~Troia G.,  Ade P.,  Bock J.,  Bond J.,  Boscaleri A.,    et~al., 2003,
  Mon.Not.Roy.Astron.Soc., 343, 284

\bibitem[\protect\citeauthoryear{{de Zotti}, {Ricci}, {Mesa}, {Silva},
  {Mazzotta}, {Toffolatti} \& {Gonz{\'a}lez-Nuevo}}{{de Zotti}
  et~al.}{2005}]{deZotti2005}
{de Zotti} G.,  {Ricci} R.,  {Mesa} D.,  {Silva} L.,  {Mazzotta} P.,
  {Toffolatti} L.,    {Gonz{\'a}lez-Nuevo} J.,  2005, Astronomy and
  Astrophysics, 431, 893

\bibitem[\protect\citeauthoryear{Elsner \& Wandelt}{Elsner \&
  Wandelt}{2009}]{Elsner2009}
Elsner F.,  Wandelt B.~D.,  2009, The Astrophysical Journal Supplement Series,
  184, 264

\bibitem[\protect\citeauthoryear{Fergusson \& Liguori}{Fergusson \&
  Liguori}{2010}]{Fergusson2010}
Fergusson J.~R.,  Liguori M.,  2010, arXiv:1006.1642, pp 1--30

\bibitem[\protect\citeauthoryear{{Freedman}, {Burns}, {Phillips}, {Wyatt},
  {Persson}, {Madore} et~al.,}{{Freedman} et~al.}{2009}]{Freedman2009}
{Freedman} W.~L.,  {Burns} C.~R.,  {Phillips} M.~M.,  {Wyatt} P.,  {Persson}
  S.~E.,  {Madore} B.~F.,    et~al., 2009, The Astrophysical Journal, 704, 1036

\bibitem[\protect\citeauthoryear{{Freedman}, {Madore}, {Gibson}, {Ferrarese},
  {Kelson}, {Sakai}, {Mould}, {Kennicutt} Jr., {Ford}, {Graham}, {Huchra},
  {Hughes}, {Illingworth}, {Macri} \& {Stetson}}{{Freedman}
  et~al.}{2001}]{Freedman2001}
{Freedman} W.~L.,  {Madore} B.~F.,  {Gibson} B.~K.,  {Ferrarese} L.,  {Kelson}
  D.~D.,  {Sakai} S.,  {Mould} J.~R.,  {Kennicutt} Jr. R.~C.,  {Ford} H.~C.,
  {Graham} J.~A.,  {Huchra} J.~P.,  {Hughes} S.~M.~G.,  {Illingworth} G.~D.,
  {Macri} L.~M.,    {Stetson} P.~B.,  2001, The Astrophysical Journal, 553, 47

\bibitem[\protect\citeauthoryear{{Gonz{\'a}lez-Nuevo}, {Toffolatti} \&
  {Arg{\"u}eso}}{{Gonz{\'a}lez-Nuevo} et~al.}{2005}]{Gonzalez2005}
{Gonz{\'a}lez-Nuevo} J.,  {Toffolatti} L.,    {Arg{\"u}eso} F.,  2005, The
  Astrophysical Journal, 621, 1

\bibitem[\protect\citeauthoryear{Gorski, Hivon, Banday, Wandelt, Hansen
  et~al.,}{Gorski et~al.}{2005}]{Gorski:2004by}
Gorski K.,  Hivon E.,  Banday A.,  Wandelt B.,  Hansen F.,    et~al., 2005,
  Astrophys.J., 622, 759

\bibitem[\protect\citeauthoryear{{Guth}}{{Guth}}{1981}]{Guth1981}
{Guth} A.~H.,  1981, Phys. Rev. D, 23, 347

\bibitem[\protect\citeauthoryear{{Guy}, {Sullivan}, {Conley}, {Regnault},
  {Astier}, {Balland}, {Basa}, {Carlberg} et~al.,}{{Guy}
  et~al.}{2010}]{Guy2010}
{Guy} J.,  {Sullivan} M.,  {Conley} A.,  {Regnault} N.,  {Astier} P.,
  {Balland} C.,  {Basa} S.,  {Carlberg} R.~G.,    et~al., 2010, Astronomy and
  Astrophysics, 523, A7

\bibitem[\protect\citeauthoryear{Hall, Keisler, Knox \& et al.}{Hall
  et~al.}{2010}]{Hall2010}
Hall N.~R.,  Keisler R.,  Knox L.,    et al. 2010, The Astrophysical Journal,
  718, 632

\bibitem[\protect\citeauthoryear{{Hicken}, {Wood-Vasey}, {Blondin}, {Challis},
  {Jha}, {Kelly}, {Rest} \& {Kirshner}}{{Hicken} et~al.}{2009}]{Hicken2009}
{Hicken} M.,  {Wood-Vasey} W.~M.,  {Blondin} S.,  {Challis} P.,  {Jha} S.,
  {Kelly} P.~L.,  {Rest} A.,    {Kirshner} R.~P.,  2009, The Astrophysical
  Journal, 700, 1097

\bibitem[\protect\citeauthoryear{Keisler, Reichardt, Aird \& et al.}{Keisler
  et~al.}{2011}]{Keisler2011}
Keisler R.,  Reichardt C.~L.,  Aird K.~A.,    et al. 2011, The Astrophysical
  Journal, 743, 28

\bibitem[\protect\citeauthoryear{Komatsu, Dunkley, Nolta, Bennett, Gold,
  Hinshaw, Jarosik, Larson, Limon, Page, Spergel, Halpern, Hill, Kogut, Meyer,
  Tucker, Weiland, Wollack \& Wright}{Komatsu et~al.}{2009}]{Komatsu2009}
Komatsu E.,  Dunkley J.,  Nolta M.~R.,  Bennett C.~L.,  Gold B.,  Hinshaw G.,
  Jarosik N.,  Larson D.,  Limon M.,  Page L.,  Spergel D.~N.,  Halpern M.,
  Hill R.~S.,  Kogut A.,  Meyer S.~S.,  Tucker G.~S.,  Weiland J.~L.,  Wollack
  E.,    Wright E.~L.,  2009, The Astrophysical Journal Supplement Series, 180,
  330

\bibitem[\protect\citeauthoryear{Komatsu, Smith, Dunkley, Bennett, Gold,
  Hinshaw, Jarosik, Larson, Nolta, Page, Spergel, Halpern, Hill, Kogut, Limon,
  Meyer, Odegard, Tucker, Weiland, Wollack \& Wright}{Komatsu
  et~al.}{2011}]{Komatsu2011}
Komatsu E.,  Smith K.~M.,  Dunkley J.,  Bennett C.~L.,  Gold B.,  Hinshaw G.,
  Jarosik N.,  Larson D.,  Nolta M.~R.,  Page L.,  Spergel D.~N.,  Halpern M.,
  Hill R.~S.,  Kogut a.,  Limon M.,  Meyer S.~S.,  Odegard N.,  Tucker G.~S.,
  Weiland J.~L.,  Wollack E.,    Wright E.~L.,  2011, The Astrophysical Journal
  Supplement Series, 192, 18

\bibitem[\protect\citeauthoryear{Komatsu \& Spergel}{Komatsu \&
  Spergel}{2001}]{Komatsu:2001rj}
Komatsu E.,  Spergel D.~N.,  2001, Phys.Rev., D63, 063002

\bibitem[\protect\citeauthoryear{Komatsu, Spergel \& Wandelt}{Komatsu
  et~al.}{2005}]{Komatsu2005}
Komatsu E.,  Spergel D.~N.,    Wandelt B.~D.,  2005, The Astrophysical Journal,
  634, 14

\bibitem[\protect\citeauthoryear{{Komatsu}, {Wandelt}, {Spergel}, {Banday} \&
  {G{\'o}rski}}{{Komatsu} et~al.}{2002}]{Komatsu2002}
{Komatsu} E.,  {Wandelt} B.~D.,  {Spergel} D.~N.,  {Banday} A.~J.,
  {G{\'o}rski} K.~M.,  2002, The Astrophysical Journal, 566, 19

\bibitem[\protect\citeauthoryear{{Kunz}, {Banday}, {Castro}, {Ferreira} \&
  {G{\'o}rski}}{{Kunz} et~al.}{2001}]{Kunz:2001ym}
{Kunz} M.,  {Banday} A.~J.,  {Castro} P.~G.,  {Ferreira} P.~G.,    {G{\'o}rski}
  K.~M.,  2001, The Astrophysical Journal, 563, L99

\bibitem[\protect\citeauthoryear{Lagache, Bavouzet, Fernandez-Conde, Ponthieu,
  Rodet, Dole, Miville-Desch\^{e}nes \& Puget}{Lagache
  et~al.}{2007}]{Lagache2007}
Lagache G.,  Bavouzet N.,  Fernandez-Conde N.,  Ponthieu N.,  Rodet T.,  Dole
  H.,  Miville-Desch\^{e}nes M.-a.,    Puget J.-L.,  2007, The Astrophysical
  Journal, 665, L89

\bibitem[\protect\citeauthoryear{Lagache \& Puget}{Lagache \&
  Puget}{2000}]{Lagache2000}
Lagache G.,  Puget J.~L.,  2000, Astronomy \& Astrophysics, 355, 17

\bibitem[\protect\citeauthoryear{Larson, Dunkley, Hinshaw, Komatsu, Nolta,
  Bennett, Gold, Halpern, Hill, Jarosik, Kogut, Limon, Meyer, Odegard, Page,
  Smith, Spergel, Tucker, Weiland, Wollack \& Wright}{Larson
  et~al.}{2011}]{Larson2011}
Larson D.,  Dunkley J.,  Hinshaw G.,  Komatsu E.,  Nolta M.~R.,  Bennett C.~L.,
   Gold B.,  Halpern M.,  Hill R.~S.,  Jarosik N.,  Kogut a.,  Limon M.,  Meyer
  S.~S.,  Odegard N.,  Page L.,  Smith K.~M.,  Spergel D.~N.,  Tucker G.~S.,
  Weiland J.~L.,  Wollack E.,    Wright E.~L.,  2011, The Astrophysical Journal
  Supplement Series, 192, 16

\bibitem[\protect\citeauthoryear{Lesgourgues}{Lesgourgues}{2011}]{Lesgourgues2011}
Lesgourgues J.,  2011, arXiv:1104.2932

\bibitem[\protect\citeauthoryear{{Lewis}}{{Lewis}}{2011}]{Lewis2011}
{Lewis} A.,  2011, Journal of Cosmology and Astroparticle Physics, 10, 26

\bibitem[\protect\citeauthoryear{Lewis, Challinor \& Lasenby}{Lewis
  et~al.}{2000}]{Lewis2000}
Lewis A.,  Challinor A.,    Lasenby A.,  2000, The Astrophysical Journal, 538,
  473

\bibitem[\protect\citeauthoryear{{Liddle} \& {Lyth}}{{Liddle} \&
  {Lyth}}{2000}]{Liddle2000}
{Liddle} A.~R.,  {Lyth} D.~H.,  2000, {Cosmological Inflation and Large-Scale
  Structure}.
Cambridge University Press

\bibitem[\protect\citeauthoryear{{Linde}}{{Linde}}{2008}]{Linde2008}
{Linde} A.,  2008, in {M.~Lemoine, J.~Martin, \& P.~Peter} ed., Inflationary
  Cosmology Vol.~738 of Lecture Notes in Physics, Berlin Springer Verlag,
  {Inflationary Cosmology}.
pp~1--+

\bibitem[\protect\citeauthoryear{{Low} \& {Tucker}}{{Low} \&
  {Tucker}}{1968}]{Low1968}
{Low} F.~J.,  {Tucker} W.~H.,  1968, Physical Review Letters, 21, 1538

\bibitem[\protect\citeauthoryear{Maldacena}{Maldacena}{2003}]{Maldacena2003}
Maldacena J.,  2003, Journal of High Energy Physics, 2003, 013

\bibitem[\protect\citeauthoryear{Matsuhara, Kawara, Sato, Taniguchi, Okuda,
  Matsumoto, Sofue, Wakamatsu \& Cowie}{Matsuhara et~al.}{2000}]{Matsuhara2000}
Matsuhara H.,  Kawara K.,  Sato Y.,  Taniguchi Y.,  Okuda H.,  Matsumoto T.,
  Sofue Y.,  Wakamatsu K.,    Cowie L.~L.,  2000, Astronomy \& Astrophysics,
  361, 407

\bibitem[\protect\citeauthoryear{Munshi, Valageas, Cooray \& Heavens}{Munshi
  et~al.}{2009}]{Munshi2009}
Munshi D.,  Valageas P.,  Cooray A.,    Heavens A.,  2009, Monthly Notices of
  the Royal Astronomical Society, 000, 1

\bibitem[\protect\citeauthoryear{{Percival}, {Reid}, {Eisenstein} \& et
  al.}{{Percival} et~al.}{2010}]{Percival2010}
{Percival} W.~J.,  {Reid} B.~A.,  {Eisenstein} D.~J.,    et al. 2010, Monthly
  Notices of the Royal Astronomical Society, 401, 2148

\bibitem[\protect\citeauthoryear{Planck-Collaboration}{Planck-Collaboration}{2011a}]{Planck-Collaboration-radioSED}
Planck-Collaboration 2011a, Astronomy \& Astrophysics, 536, A15

\bibitem[\protect\citeauthoryear{Planck-Collaboration}{Planck-Collaboration}{2011b}]{Planck-Collaboration-ERCSC}
Planck-Collaboration 2011b, Astronomy \& Astrophysics, 536, A7

\bibitem[\protect\citeauthoryear{Planck-Collaboration}{Planck-Collaboration}{2011c}]{Planck-Collaboration-statradio}
Planck-Collaboration 2011c, Astronomy \& Astrophysics, 536, A13

\bibitem[\protect\citeauthoryear{Planck-Collaboration}{Planck-Collaboration}{2011d}]{Planck-Collaboration-CIB}
Planck-Collaboration 2011d, Astronomy \& Astrophysics, 536, A18

\bibitem[\protect\citeauthoryear{Planck-Collaboration}{Planck-Collaboration}{2011e}]{Planck-Collaboration-Dust}
Planck-Collaboration 2011e, Astronomy \& Astrophysics, 536, A24

\bibitem[\protect\citeauthoryear{Puget, Abergel, Bernard, Boulanger, Burton,
  Desert \& Hartmann}{Puget et~al.}{1996}]{Puget1996}
Puget J.-L.,  Abergel A.,  Bernard J.-P.,  Boulanger F.,  Burton W.~B.,  Desert
  F.-X.,    Hartmann D.,  1996, Astronomy \& Astrophysics, 308, 5

\bibitem[\protect\citeauthoryear{{Renaux-Petel}}{{Renaux-Petel}}{2009}]{Renaux-Petel2009}
{Renaux-Petel} S.,  2009, Journal of Cosmology and Astroparticle Physics, 10,
  12

\bibitem[\protect\citeauthoryear{{Riess}, {Macri}, {Casertano}, {Sosey},
  {Lampeitl}, {Ferguson}, {Filippenko}, {Jha}, {Li}, {Chornock} \&
  {Sarkar}}{{Riess} et~al.}{2009}]{Riess2009}
{Riess} A.~G.,  {Macri} L.,  {Casertano} S.,  {Sosey} M.,  {Lampeitl} H.,
  {Ferguson} H.~C.,  {Filippenko} A.~V.,  {Jha} S.~W.,  {Li} W.,  {Chornock}
  R.,    {Sarkar} D.,  2009, The Astrophysical Journal, 699, 539

\bibitem[\protect\citeauthoryear{Righi, Hern\'{a}ndez-Monteagudo \&
  Sunyaev}{Righi et~al.}{2008}]{Righi2008}
Righi M.,  Hern\'{a}ndez-Monteagudo C.,    Sunyaev R.~A.,  2008, Astronomy and
  Astrophysics, 478, 685

\bibitem[\protect\citeauthoryear{{Sachs} \& {Wolfe}}{{Sachs} \&
  {Wolfe}}{1967}]{SW1967}
{Sachs} R.~K.,  {Wolfe} A.~M.,  1967, The Astrophysical Journal, 147, 73

\bibitem[\protect\citeauthoryear{{Sajina}, {Partridge}, {Evans}, {Stefl},
  {Vechik}, {Myers}, {Dicker} \& {Korngut}}{{Sajina} et~al.}{2011}]{Sajina2011}
{Sajina} A.,  {Partridge} B.,  {Evans} T.,  {Stefl} S.,  {Vechik} N.,  {Myers}
  S.,  {Dicker} S.,    {Korngut} P.,  2011, The Astrophysical Journal, 732, 45

\bibitem[\protect\citeauthoryear{Sehgal, Bode, Das, Hernandez-Monteagudo,
  Huffenberger, Lin, Ostriker \& Trac}{Sehgal et~al.}{2010}]{Sehgal2010}
Sehgal N.,  Bode P.,  Das S.,  Hernandez-Monteagudo C.,  Huffenberger K.,  Lin
  Y.-T.,  Ostriker J.~P.,    Trac H.,  2010, The Astrophysical Journal, 709,
  920

\bibitem[\protect\citeauthoryear{Seljak \& Zaldarriaga}{Seljak \&
  Zaldarriaga}{1996}]{Seljak1996}
Seljak U.,  Zaldarriaga M.,  1996, The Astrophysical Journal, 469, 437

\bibitem[\protect\citeauthoryear{Serra \& Cooray}{Serra \&
  Cooray}{2008}]{Serra2008}
Serra P.,  Cooray A.,  2008, Physical Review D, 77, 1

\bibitem[\protect\citeauthoryear{Smoot, Bennett, Kogut \& et al.}{Smoot
  et~al.}{1992}]{Smoot1992}
Smoot G.~F.,  Bennett C.~L.,  Kogut A.,    et al. 1992, The Astrophysical
  Journal, 396, L1

\bibitem[\protect\citeauthoryear{Spergel \& Goldberg}{Spergel \&
  Goldberg}{1999}]{Spergel1999}
Spergel D.,  Goldberg D.,  1999, Physical Review D, 59, 1

\bibitem[\protect\citeauthoryear{{Starobinski{\v i}}}{{Starobinski{\v
  i}}}{1979}]{Starobinski1979}
{Starobinski{\v i}} A.~A.,  1979, Soviet Journal of Experimental and
  Theoretical Physics Letters, 30, 682

\bibitem[\protect\citeauthoryear{{Sunyaev} \& {Zeldovich}}{{Sunyaev} \&
  {Zeldovich}}{1972}]{Sunyaev1972}
{Sunyaev} R.~A.,  {Zeldovich} Y.~B.,  1972, Comments on Astrophysics and Space
  Physics, 4, 173

\bibitem[\protect\citeauthoryear{{Toffolatti}, {Argueso Gomez}, {de Zotti},
  {Mazzei}, {Franceschini}, {Danese} \& {Burigana}}{{Toffolatti}
  et~al.}{1998}]{Toffolatti1998}
{Toffolatti} L.,  {Argueso Gomez} F.,  {de Zotti} G.,  {Mazzei} P.,
  {Franceschini} A.,  {Danese} L.,    {Burigana} C.,  1998, Monthly Notices of
  the Royal Astronomical Society, 297, 117

\bibitem[\protect\citeauthoryear{Viero, Ade, Bock \& et al.}{Viero
  et~al.}{2009}]{Viero2009}
Viero M.~P.,  Ade P. a.~R.,  Bock J.~J.,    et al. 2009, The Astrophysical
  Journal, 707, 1766

\bibitem[\protect\citeauthoryear{Wick}{Wick}{1950}]{Wick1950}
Wick G.~C.,  1950, Phys. Rev., 80, 268

\end{thebibliography}

\label{lastpage}

\end{document}